# Nanoscale Heat Transfer – from Computation to Experiment

Tengfei Luo [a] and Gang Chen*[b]

Heat transfer can differ distinctly at the nanoscale from that at the macroscale. Recent advancement in computational and experimental techniques has enabled a large number of interesting observations and understanding of heat transfer processes at the nanoscale. In this review, we will first discuss recent advances in computational and experimental methods used in nanoscale thermal transport studies, followed by reviews of novel thermal transport phenomena at the nanoscale observed in both computational and experimental studies, and discussion on current understanding of these novel phenomena. Our perspectives on challenges and opportunities on computational and experimental methods are also presented.

## Introduction

Heat transfer at the nanoscale can differ distinctly from that predicted by classical laws.[1] Understanding nanoscale heat transfer will help thermal management of electronic, optical, and optoelectronic devices, and design new materials with different thermal transport properties for energy conversion and utilization. Research on nanoscale heat transfer has advanced significantly over the past two decades, and a large number of interesting phenomena have been observed.

The record-high thermal conductivities were computationally calculated (6600 W/mK at room temperature)[2] and experimentally measured (3500 W/mK at room temperature)[3] in a single-walled carbon nanotube (CNT), a one-dimensional system with thermal conductivity exceeding that of diamond – the best known heat conductor in bulk form. The thermal conductivities of CNTs are not only high but could also be diverging due to the one-dimensional nature of thermal transport.[4-6] Graphene[7] has also been reported to have exceptionally high thermal conductivity (5800 W/mK).[8] Polymers, such as polyethylene, are usually regarded as thermal insulators in the amorphous phase (~0.3 W/mK).[9] However, simulations suggest that along a polyethylene molecular chain, thermal conductivity is high and could even be divergent.[10] Ultra-drawn polyethylene nanofibers are found to have thermal conductivities (~100 W/mK) hundreds of times higher than their amorphous counterparts.[11] In fluids suspended with nanoparticles, thermal conductivity enhancement way beyond the amount predicted by traditional effective medium theory has been reported.[12, 13] This observation has inspired a large amount of research both theoretically and experimentally to explore the mechanism and applications of the enhanced thermal transport in nanofluids.[14-19] Theories have predicted that radiative heat transfer between objects separated at small distances increase significantly with decreasing separations, supported by a few prior experiments performed at micrometer or larger separations.[20-22] Recent experiments have pushed the separations between surfaces to tens of nanometers, and radiative heat transfer many orders of magnitude higher than that predicted by Planck's blackbody radiation law has been reported.[23]

Besides the high and increased thermal conductivity, low and reduced thermal conductivities in materials have also been observed and achieved. Superlattices have thermal conductivities way below the predictions based on the Fourier law,[24-26] and even below that of alloys,[27, 28] and what the minimum thermal conductivity theory predicts.[29] Significantly reduced thermal conductivity has also been found in different nanowires, especially silicon,[30-32] some of which are beyond what current theories can explain. $WSe_2$, even in a crystal structure, has been found to have thermal conductivity only twice of that of air.[33] Such observations have stimulated research on thermoelectric energy conversion, and understanding derived from the studies is applied to the synthesis of bulk nanostructured thermoelectric materials.[34]

Many of these measured exceptional thermal conductivities have been explained by computational and experimental studies, and many significant questions remain. In this article, we will first review recent advances in computational and experimental methods in nanoscale heat transfer, and then discuss how they are used to understand and explain nanoscale thermal transport phenomena. We will also discuss unsolved problems and challenges. Due to the breadth of the field, we cannot cover all topics but choose to focus on those that we are most familiar with. Readers interested in this field should also consult other reviews and textbooks.[1, 35, 36]

## I. Advances in Computational and Experimental Methods

In this section, we will give a summary of some computational and experimental tools used for simulating and experimentally probing nanoscale heat transfer. Most efforts in the past have been devoted to understand heat conduction in nanostructured materials, especially crystalline-based nanostructures, for which transport can be adequately described by phonons – quantized lattice waves. Progress has been made in both computational and experimental tools to understand phonon transport.

### I. 1 Computational Methods

Thermal conductivities of crystals are often modeled following the strategies developed by Klemens,[37] Callaway[38] in the 1950s based on phonon Boltzmann equation developed by Peierls,[39] Debye approximation for phonon dispersion, and analytical expressions derived on phonon scattering rates using the Fermi golden rule with fitting parameters. These fitting parameters are often determined by matching measured temperature dependence of thermal conductivity with kinetic expressions of the thermal conductivity integral. Although such an approach has led to qualitative understanding of phonon transport, the extracted details from such models may be very inaccurate. These deficiencies are better seen when the same information are applied to explain experimental results in nanostructures. Recently, there has been significant progress in simulation tools, as summarized below.

Computational methods including lattice dynamics based on force constants from empirical force fields or first-principles calculations, molecular dynamics simulations, Green's function, Boltzmann transport equation and Monte Carlo simulations have been used to study nanoscale thermal transport. This section introduces basics of these methods and reviews recent advances of them in nanoscale heat transfer.

For heat conduction by phonons, a widely used expression to evaluate thermal conductivity is from the kinetic theory based on the single-mode relaxation time approximate:[1]

$$\kappa = \frac{1}{V} \sum_{\vec{k}\lambda} c_{\vec{k}\lambda} v_{\vec{k}\lambda}^2 \tau_{\vec{k}\lambda} = \frac{1}{V} \sum_{\vec{k}\lambda} c_{\vec{k}\lambda} v_{\vec{k}\lambda} \Lambda_{\vec{k}\lambda} \qquad (1)$$

where $\kappa$ is thermal conductivity, $c$ the heat capacity, v the group velocity, $\tau$ the relaxation time of phonon ($\vec{k}$, $\lambda$) with rest of phonons, and $\Lambda$ the phonon mean free path ($\Lambda = v \cdot \tau$). The summation is over all the phonon modes in the first Brillouin zone, with the subscript $\vec{k}$ referring to the wave-vector and $\lambda$ referring to different phonon branches.

The phonon properties ($c$, v, $\tau$) included in Eq. (1) can be obtained through lattice dynamics and molecular dynamics simulations. For a material, the total lattice energy can be expressed by a Taylor series expansion with respect to atomic displacements,

$$U = U_o + \frac{1}{2!}\sum_{ij} \Phi_{ij} u_i u_j + \frac{1}{3!}\sum_{ijk} \Psi_{ijk} u_i u_j u_k + O[u^4] \qquad (2)$$

where $\Phi$ and $\Psi$ are the harmonic and cubic force constants. Using the harmonic force constants, one can calculate the phonon dispersion relation, from which the group velocity and specific heat can be obtained. Impacts of the cubic term on specific heat and phonon group velocity are usually small. However, it dominates phonon-phonon scattering, and the cubic force constants can be used to evaluate phonon-phonon relaxation times using Fermi's gold rule.

Before computational methods are developed to obtain accurate force constants, the phonon properties are usually obtained experimentally or through simplified analytical models developed by Klemens,[40] Holland,[41] Slack[42] and Callaway,[38] mostly based on the Debye approximation for the phonon dispersion and need fitting parameters. In recent years, lattice dynamics with inputs from atomistic models, especially first-principles density functional theory (DFT) calculations, has significantly improved the accuracy of predicted phonon properties and thermal conductivity of materials, especially crystals.[43-45] These methods do not involve fitting to experimental data.

### I. 1.1 First-Principles Calculations

Evaluating force constants using first-principles calculations has been made possible thanks to the aggressive improvements in computational hardware and software. When combined with first-principles calculations, lattice dynamics has been able to predict phonon properties and thermal conductivity with unprecedented accuracy and without the need of any empirical input.[43-45] In the early 1980s, Yin and Cohen successfully used first-principles calculations combining a finite difference method to calculate the phonon frequencies and mode Grüneisen parameters at a few high symmetry points in the first Brillouin zone for Si and Ge.[46] This method, however, involves the evaluation of a large set of energy-displacement data to calculate the derivatives included in Eq. (2). Each energy-displacement datum requires a first-principles calculation. For information on high symmetry points in the Brillouin zone, primitive cell calculations, which involve the minimum amount of atoms, are enough. However, when phonon properties of arbitrary phonon modes in the Brillouin zone are needed, supercells several times larger than the primitive cell are needed so that the real space information can be transformed into the points inside the first Brillouin zone. Since the first-principles calculations on large systems are much more computationally expensive, these calculations can easily become overwhelming.

Giannozzi et al.[47] used a linear-response approach within the DFT framework to obtain the harmonic force constants which is then used to calculate phonon dispersion. Their prediction agreed very well with experimental data. The theoretical basics of such a method is that the harmonic force constants of crystals are determined by their static linear electronic response.[48, 49] The small coordinate variations of the nuclei from their equilibrium positions, which correspond to phonons, work as static perturbations to the electron system. Such a method, called the density functional perturbation theory, calculates the matrix of the force constants in the reciprocal space directly and thus only needs primitive cell calculations, avoiding the expensive supercell calculations.

Calculation of cubic force constants are even more time consuming when using the finite difference method since a larger energy-displacement data needs to be obtained to evaluate the second derivative in Eq. (2). Gonze and Vigneron[50] proposed a general scheme using the "2n+1" theorem[51-53] of the perturbation theory within DFT for an accurate computation of a large class of nonlinear-response coefficients of crystalline solids. This density functional perturbation theory method also uses primitive cell calculations. The 2n+1 theorem enables the calculation of the third-order derivatives of the lattice energy by using only a first-order perturbation calculation, significantly lowering the computational cost. Debernardi et al.[54, 55] used this method to obtain the cubic force constants for several crystals and calculated the phonon lifetimes at the Brillouin zone center. Excellent agreement between their calculated data and Raman spectroscopy data was found.[31] However, the information at Brillouin zone center is not enough for thermal conductivity evaluation, which need phonon properties over the whole first Brillouin zone as shown in Eq. (1).

To calculate the thermal conductivity, Broido and coworkers[56] used the density functional perturbation theory method to calculate the harmonic and cubic force constants to obtain dynamical matrices over a set of discrete points in the first Brillouin zone. Based on these parameters, an exact solution of a linearized phonon Boltzmann equation was used to calculate the thermal conductivity of crystalline silicon, germanium[56] and diamond.[43] Their predictions are within 10% of experimental values. Such a prediction power surpasses all the previous calculations based on phenomenological models and empirical interatomic potentials. The key factor to the success of this approach is the use of accurate force constants from density functional perturbation theory calculations. However, for crystals with complicated primitive cells, such as $Bi_2Te_3$ and GaN, the computational cost of evaluating the cubic force constants using density functional perturbation theory becomes prohibitively high. Esfarjani and Stokes[57] used a method to obtain the force constants by fitting them to a set of force-displacement data. The forces on atoms can be calculated using DFT which greatly reduces the computational cost. Esfarjani et al.[45] used this method to calculate the thermal conductivity of silicon and showed that the accuracy is comparable to Broido's calculation.[56] This method also enabled the prediction of thermal conductivity of more complicated materials such as GaAs,[58] half-heusler,[59] PbTe[60] and PbSe.[61] These results as functions of temperature together with experimental values are summarized in Fig. 1. Good agreement between the calculated thermal conductivity and experimental data can be seen.

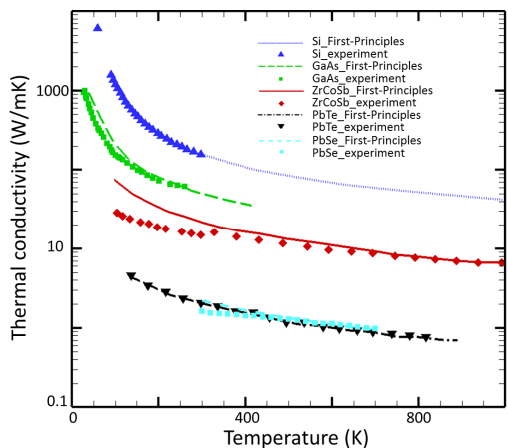

Figure 1 Thermal conductivity of Si, GaAs, ZrCoSb (half-heulser), PbTe, PbSe, AlAs and their comparison with experiments. (Si – first-principles calculation[62] and experiment[63]; GaAs – first-principles calculation[58] and experiment[64]; ZrCoSb – first-principles calculation[59] and experiment[65, 66]; PbTe – first-principles calculation[67] and experiment[68]; PbSe – first-principles calculation[69] and experiment[70].)

The first-principles lattice dynamics method has also been applied by Garg et al.[71] to alloys to study the role of disorder and anharmonicity in thermal conductivities, and excellent agreement with experimental data has been obtained. This method has also been applied to short-period superlattices[72] to study how the thermal conductance can be enhanced by carefully engineering the mass mismatch between the constituent materials in the superlattice.

Unlike first-principles methods, empirical potential functions use analytical forms to describe the lattice energy as a function of the atomic coordinates. As a result, the force constants in Eq. (2) can be obtained analytically, which is computationally efficient. However, empirical potentials can be inaccurate for thermal transport studies since they are usually not designed for such purposes. First-principles calculations can be used to optimize the empirical potentials to improve their accuracy in predicting phonon properties. Such a scheme has been successfully used by Broido and coworkers to predict the thermal conductivity and study the thermal transport physics of a wide variety of materials including CNT,[73] graphene,[74] graphite,[75] and boron nitride.[76]

It is worth noting that all the aforementioned methods, except those used by Broido et al.,[5, 74, 76] use the single mode phonon relaxation time approximation which includes both normal and Umklapp scatterings as sources of thermal resistance. The relaxation time approximation has been proven accurate for most of the cases involving three dimensional crystals where Umklapp scattering dominates. However, it has been pointed out that in low dimensional materials, such as CNT and graphene, the relaxation time approximation can significantly underestimate thermal conductivity due to the significance of normal phonon-phonon scattering process.[5, 74] For some three dimensional crystals like diamond,[43] in which normal scattering is also important, the relaxation time approximation also underestimates thermal conductivity.

### I. 1.2 Classical Molecular Dynamics Simulations

Classical molecular dynamics simulations trace the time-dependent trajectories of all atoms in the simulation domain based on the Newton's second law of motion and interatomic potentials. Nonequilibrium molecular dynamics (NEMD) and equilibrium molecular dynamics (EMD) simulations are the two major methods to calculate thermal transport properties like thermal conductivity and interfacial thermal conductance. These methods are discussed in details in many references.[2, 77-84] Molecular dynamics has also been developed to study detailed phonon properties such as density of state, dispersion relation, relaxation time and transimssion across interfaces. Here, we discuss a few studies to show the evolving capability of molecular dynamics in studying thermal transport.

Phonon relaxation time is a central parameter in thermal transport in nanostructures.[85] Li and Yip[86] used exponential functions to fit the heat current autocorrelation function to obtain a time constant for the overall thermal relaxation. Volz and Chen[82] also derived a formula with which the averaged phonon relaxation time can be extracted from the heat flux autocorrelation function. These methods and their derivatives have been used to obtain averaged phonon relaxation times of other materials.[87-89] Later, McGaughey and Kaviany[90] successfully extracted the relaxation time of arbitrary phonon mode in a crystal through molecular dynamics simulations by analyzing the decay of the energy autocorrelation function of the specific phonon mode. This was made possible through tracking the mode energy by projecting atomic trajectory onto phonon eigenvectors. Using a similar method, Henry and Chen[10, 91, 92] studied phonon transport in silicon and single polyethylene chains. An alternative to analyzing autocorrelation functions to extract phonon relaxation times has been developed by McGaughey and coworkers[93] by calculating the line-width of spectral energy density of specific phonon modes. This method was first applied to CNTs,[93] and later used to study PbTe by Qiu et al.[94] and ZrCoSb by Shiomi and coworkers.[59]

The drawbacks of molecular dynamics are the availability, reliability and transferability of the empirical potentials. One solution is using first-principles calculations to evaluate atomic forces. This method has been applied to system with small mean free path to predict their thermal conductivity.[95] It has also been used by Luo and Lloyd to study thermal transport across ultrathin superlattices.[96] However, such quantum molecular dynamics approach is not computationally efficient to handle large systems. Building empirical potential functions by fitting to DFT calculations has been increasingly used, especially for complicated and unconventional materials where potentials are not available. Such an approach has been successfully used by several groups to study thermal transport in metal-organic framework,[97] PbTe[94] and $Bi_2Te_3$.[98, 99] Instead of using empirical functions, Esfarjani and coworkers[45, 59] used the force constants obtained from first-principles to calculate the interatomic forces for molecular dynamics simulations according to Eq. (2). Good accuracy of phonon properties and thermal conductivities has been obtained for silicon[45] and ZrCoSb.[59]

According to the above discussion, potential models are central factors that determine the accuracy of thermal conductivity prediction. To demonstrate this, thermal conductivity of silicon calculated by different methods and potentials are summarized in Fig. 2. It is seen that first-principles methods are generally much more accurate than empirical potential methods. Among different potentials models, large differences in predicted thermal conductivity also exist. It is also found that traditional potentials like Tersoff potential can be optimized to better produce thermal conductivity data.

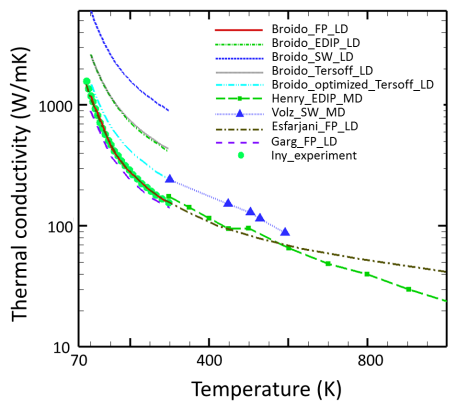

Figure 2 Silicon thermal conductivities from calculations and their comparison with experimental data, highlighting the importance of accurate potential on predicting thermal conductivity (FP- first-principles; LD-lattice dynamics; Broido data;[56, 100] Henry data;[91] Volz data;[82] Esfarjani data;[62] Garg data;[101] Inyushkin data.[63] Note: Esfarjani_FP_LD data overlap with Broido_FP_LD data; Broido_EDIP_LD data almost overlap with Broido_Tersoff_LD data.)

The above-mentioned methods, including first-principles and molecular dynamics, rely on tracking specific phonon modes, either through Fermi's golden rule or through atomic trajectory, to obtain the phonon relaxation times. These methods have been proven successful for studying phonon relaxation times due to phonon-phonon scatterings and isotope scattering. However, these methods have not been able to accurately include scatterings due to large scale defects such as dislocations, grain boundaries and interfaces. Molecular dynamics is promising in tackling these problems if accurate potential models are available to describe the realistic structure of dislocations, grain boundaries and interfaces.

At interfaces between materials, reflections of phonons cause thermal boundary resistance. Molecular dynamics simulations have been applied to study phonon transport across interfaces. Nonequilibrium molecular dynamics are often used to compute thermal boundary resistances. To understand better thermal boundary resistance, however, one needs pictures of phonon reflection and transmission at interfaces. Keblinski and coworkers[102] studied detailed phonon transmission at interfaces by launching a series of phonon wave-packets towards the interface one at a time in a molecular dynamics simulation. The energies of these phonon wave-packets are tracked before and after they hit the interface, and the transmission coefficient is determined as the ratio of the energy of the transmitted wave-packet and that of the incident wave-packet. This method has later been applied to study phonon transmission at a number of different interfaces.[103-107] Such a phonon wavepacket method, however, is limited to studying a single phonon mode at near zero temperature where phonon-phonon interactions are absent, and to phonon incident angles close to normal to the interface due to limitations of the simulation domain size.

Other methods such as analyzing the overlaps of the vibration power spectra of two contacting materials[96, 108-116] and calculating the spectral temperatures of vibrations with different frequencies[114, 117] have also been used to qualitatively study harmonic and anharmonic thermal transport across interfaces. Quantitative analyses using these tools still need to be developed.

Overall, a major challenge in molecular dynamics simulations lies in the accuracy of the potential models. For crystalline materials, this can be overcome by using potentials optimized by first-principles calculations or use forces calculated from first-principles directly. However, systems with interfaces and defects cannot be easily simulated by first-principles. One might use potential functions that are able to model chemical reactions to correctly describe interface and defect structures.[115] Such kinds of potentials, however, are very rare and are only applicable to a limited number of materials.[118, 119] The generalization of these potential models will be important for molecular dynamics simulations of nanoscale thermal transport.

**I. 1.3 Atomistic Green's Function**

Atomistic Green's function has emerged as a useful approach to study thermal transport across interfaces. A Green's function is a mathematical method for solving differential equations such as the Schrodinger equation. In an atomistic Green's function approach, the system is represented at the molecular level by atomistic potential models. Heat current in the system subject to a small temperature difference is related to the interatomic force constants. This heat current is expressed in terms of Green's function, and the phonon transmission as a function of phonon frequency is calculated. Mingo and Yang[120] used a non-equilibrium harmonic atomistic Green's function to study the interfacial heat transfer between a nanowire and an amorphous oxide overlayer using the Bethe lattice approach[121] with a self-consistent flux model. This model included realistic silicon phonon dispersion and was able to predict the transmission coefficients as a function of phonon frequencies and interface contact strengths. Anharmonicity has later been incorporated by the same authors to study the heat flow through molecular junctions.[122] Atomistic Green's function has also been used to study a range of structures from one dimensional model systems[123, 124] to more realistic systems such as Ge-Si/Ge superlattice[125] and graphene-metal interfaces.[126]

However, not all of these calculations include anharmonicity, whose role in interfacial thermal transport is not currently clear. Mingo and Yang found that the anharmonic effect is negligible for silicon nanowire with characteristic lengths smaller than 20nm at room temperature.[120] This effect, however, is material dependent and temperature dependent. The interatomic interactions used in the past atomistic Green's function calculations are mainly from empirical models. Although Fisher and coworkers[126] used DFT to model the interaction of bulk materials, the interfacial interaction still remains empirical. These calculations studied perfect interfaces without considering detailed interface structures, such as the inter-diffusion of atoms and surface reconstruction which exist in realistic interfaces. Accurate interatomic interaction models, especially those are able to correctly model the interfacial microstructure, need to be developed. Without the realistic description of the interface structure, it is difficult for atomistic Green's function to provide accurate phonon transmission coefficients. Usually, Green's functions calculate transmittance as a function of phonon frequency. Phonon mode specific transmittance is also important, and such information from the atomistic Green's function approach has been recently reported by Huang et al.[127] on graphene and graphene nano ribbon systems.

**I. 1.4 Multiscale Simulation**

The DFT and molecular dynamics simulation methods discussed above are limited to small simulation domain. Periodic boundary conditions often used in these simulations make it routine to simulate bulk crystals, although care should be paid in molecular dynamics simulations to avoid additional correlations caused by the periodic boundary conditions. Multiscale simulation tools are needed to bridge the length-scale and time-scale gaps between atomistic simulations and real materials and devices. With information obtained on phonon relaxation time in bulk materials and interfacial transmission, phonon Boltzmann equation has been an important multiscale tool for thermal transport studies. The phonon distribution function in Boltzmann equation can be a function of position, direction, time, frequency, wave vector as well as polarization,[128] making solving phonon Boltzmann equation challenging. Different methods, such as the gray model, semi-gray model and non-gray model, exist to simplify the solution of Boltzmann equation to different extends. These approximations are discussed in details in different texts[1] and review articles.[129]

The relaxation time approximation is usually necessary to make solving Boltzmann equation computationally feasible. To obtain the relaxation times, simplified models[38, 40-42] with adjustable parameters determined by fitting to experimental data are usually used. These approaches, however, are not always accurate. On the other hand, the development of molecular dynamics and first-principles calculations has made more accurate phonon relaxation times available.[45, 91] These values have been used as input in Boltzmann equation modeling.[130] Exact solutions of phonon Boltzmann equation, which can yield more precise prediction than relaxation time approximation, are much more computationally expensive to obtain.

For arbitrary structures, Monte Carlo simulations are often used to solve Boltzmann equation.[131-136] However, when frequency-dependent phonon mean free paths need to be considered, the computational time becomes very large.[132] Recently developed Monte Carlo schemes using a control-variate variance-reduction formulation for obtaining solutions of phonon Boltzmann equation has been proven to speedup traditional Monte Carlo methods by nine orders of magnitude.[137] The authors also showed that additional computational benefit is possible in the limit that the governing equation can be linearized.[138] These make more detailed thermal transport simulation for complicated nanostructures within reach. Instead of using a fixed mean free path in Boltzmann equation, McGaughey and Jain used a Monte Carlo scheme to sample the free path of phonon-phonon scattering with a Poisson distribution and added boundary scattering by launching phonons in a random position in the structure modelled in the simulations.[139] A similar method was used by Sofo and Mahan to study carrier transport in synthetic opals for thermoelectrics applications.[140] The differences between McGaughey and Jain's Monte Carlo simulation and the widely used Matthiessen rule were found to be large especially when the structure sizes have a large variation. A recent study by Peraud and Hadjiconstantinou,[138] however, found that such a treatment can significantly over estimate thermal conductivity when applied to structures more complex than a thin film.

Another important multiscale simulation method is the effective media approximation.[141] It has long been used to predict properties of composite materials. Hasselman and Johnson[142] and Benveniste[143] derived theoretical models to account for thermal boundary resistance. For macrocomposites where macroscopic thermal properties can be used, effective media approximation is usually accurate. However, in nanocomposites, where interface scattering becomes important since composite phases can have sizes smaller than phonon mean free path, the use of macroscopic thermal conductivity will result in inaccurate results.[132, 144] A modified effective media approximation using adjusted thermal conductivity according the classical size effect has been proposed, and much improved results have been obtained.[145] The reduced thermal conductivity of nanostructures and interfacial thermal resistance obtained from atomistic simulations can also be used as inputs for effective media approximation to enable accurate thermal conductivity prediction for nanocomposites. Existing effective media theories are usually limited to situations when one phase is dilute. However, when the concentration of the minor phase increases to a limit where particle percolation occurs, thermal boundary resistance between the contacting particles also needs to be included besides the particles-matrix resistance.

One major challenge exist in multiscale simulation is correctly simulating interfacial phonon transmission at interfaces – an important component in many nanostructured materials such as superlattices and nanocomposites. Although transmission coefficients can potentially be obtained from atomistic simulations such as molecular dynamics and atomistic Green's function, the aforementioned limitations of these methods make obtaining accurate transmission coefficients difficult.

**I. 2 Experimental Methods**

A range of experimental methods and apparatus has been optimized and developed to measure thermal transport properties and study fundamental heat transfer physics. These methods have enabled the observation of many interesting thermal transport phenomena at the nanoscale. In this section, these important experimental methods are briefly reviewed, and perspectives on some of them are discussed.

**I. 2.1 $3\omega$ Method**

One of the most important methods to measure thermal conductivity is the $3\omega$ method developed by Cahill and co-workers.[146, 147] The name $3\omega$ derives from the fact that when a resistor is heated with a sinusoidal wave at frequency $\omega$, its temperature rise is proportional to $2\omega$, and there is a third harmonic in voltage drop in the resistor due to this $2\omega$ temperature rise. By measuring this $3\omega$ voltage, one can determine the temperature of the resistor. Originally, this method was developed to measure thermal conductivity of materials at high temperatures to reduce radiation heat loss.[148] Narrow resistors can be lithographically patterned on the substrate or thin film. For thin-film thermal conductivity measurement, narrow heaters can generate a high local heat flux such that temperature drop across the film is comparable to that in the substrate on which the film is deposited on. This method and its derivatives have been applied to a large variety of materials including solids,[148] thin films,[149, 150] wires and CNT bundles,[151] individual CNT,[152] liquids,[147, 153] nanoscale junctions,[154] and nanotubes arrays.[155] For thin-films on a substrate, $3\omega$ method usually measure thermal conductivity in the direction perpendicular to the film plane. By using heaters with width comparable to the film thickness to explore heat spreading inside the film, however, both in-plane and cross-plane thermal conductivity can be determined.[156] Dames and Chen extended the $3\omega$ method by applying a DC component in the driving current to the heater so that the thermal conductivity can also be extracted by lower order harmonics at $1\omega$ and $2\omega$.[157] Depending on the property of the system, harmonics on one order can be chosen over the others to produce better results.

For in-plane thermal conductivity measurement,[158] a key challenge is that heat leakage through the substrate. One strategy is to remove the substrate using methods like chemical etching[159, 160] and plasma etching.[161] Using an insulator between the film and the substrate, one can also force heat conduction along the film.[162]

**I. 2.2 Nanowire and Nanotube Measurement Platform**

Nanowires and nanotubes have small cross-sections and consequently small amount of heat flow. Thermal conductivity measurements through nanowires and nanotubes require methods to measure such small heat flow. To measure thermal conductivity of multi-walled CNTs, Kim et al.[163] developed a microfabricated measurement platform with a CNT embedded in it. This measurement platform has been improved by Shi et al.[164] and applied to different nanostructures like single walled CNT,[165] nanowires,[30, 166, 167] suspended and supported graphene,[168] and nanocontacts.[169, 170] This category of platforms has enabled the measurements of many novel thermal transport properties, and some of them are discussed in later sections. Other microfabricated structures have also been developed to measure thermal conductivity of nanotubes.[152, 171, 172] Dames et al.[173] used Wollaston wire to eliminate the fabrication and a special sample holder to measure thermal conductivity of a carbon nanotube inside a transmission electron microscope.

**I. 2.3 Scanning Thermal Microscopy**

High spatial-resolution temperature measurement is desirable in many applications. Optical methods are limited by the diffraction limit. Contact measurements combining high spatial resolution temperature sensors with scanning probe microscope has been pursued by various groups. In fact, the first scanning thermal microscope was invented by Williams and Wickramasinghe[174] before atomic force microscope was invented by Binnig et al.[175] to overcome the limit of scanning tunneling microscope for topology measurements of conducting samples only. In Wickramasinghe's scanning thermal microscope, heat conduction between the scanning thermal microscope and a substrate is used to infer the surface topology of non-electrical conducting samples. Majumdar[171, 174] replaced the atomic force microscope tip with a thermocouple probe for contact mode spatially resolved temperature measurement. Resistive thermometer made from Wollaston wires were used in some commercial scanning thermal microscope.[176] Microfabricated cantilevers with thermosensors at the tip were later developed to improve the spatial resolution.[177-180] These scanning thermal microscopes have been used for a wide variety of applications such as measuring surface topology,[181] detecting phase changes,[182] and measuring thermal conductivity.[183]

Typically, scanning thermal microscope has a lateral spatial resolution of about 100nm and a temperature resolution of 1mK.[184] However, with nanofabricated tips, a spatial resolution of about 10nm has been reported.[178] Recently, experiment performed in ultra-high vacuum has been capable to achieve temperature resolution of ~15mK and spatial resolution of ~10nm.[185] To eliminate thermal transport due to air between the cantilever and the sample surface which influences the quantitative data analysis, double scanning technique[186] and experiments in vacuum[187] have been employed. One of the major challenges of scanning thermal microscopy is how to separate thermal data from topographic data since the heat flux required to maintain the tip temperature depends on the thermal properties of the sample.[188] Various methods have been developed to overcome this challenge such as replying on the thermal expansion,[189, 190] probes with very low thermal conductance,[191] specially fabricated probes shaped to probe and cantilever at the same time.[180, 188] Menges et al.[187] recently used a two-step method involving determining local temperature from the difference between the measured heat flux for heat sources switched on and off, and a quantitative measurement of thermal conductance and temperature distribution with nanoscale resolution has been achieved. Despite these efforts and progress made, the coupling between topology and heat transfer continues to plague the wide application of scanning thermal microscope.

### I. 2.4 Bi-Material Cantilever

Cantilevers made of two material layers with different thermal expansion coefficients can work as thermal sensors to measure temperature and heat flow. These cantilevers have been used in many applications such as scanning microscopy probes,[184, 192] IR detector[193] and thermal actuator.[194] With the right combination of materials, miniaturized bi-material cantilevers were able to detect a power below 100pW due to the small thermal mass.[195] The resolution has been recently further improved to below 4pW.[196] Such high resolutions enable their application in measuring very small thermal signals. The very high sensitivity and resolution has recently enabled the observation of remarkable thermal radiation at near-field[23] and the measurement of ultrahigh thermal conductivity of polymer nanofibers.[11]

### I. 2.5 Optical Methods

Optical methods have been developed for thermal property measurements of thin-films. Using optical methods, it is usually difficult to do absolute thermal conductivity measurements because optical heat input to the sample is hard to measure accurately. Time variations are often used to obtained thermal diffusivity ($k/\rho c$, where k is thermal conductivity, $\rho$ density and c specific heat) or effusivity ($k\rho c$). Both period heating and transient heating have been used. In periodic heating methods, temperature rise at the surface causes either reflectance change,[197-199] emittance change,[200-203] absorptance change,[204] photoacoustic signal change,[205] thermal expansion on the sample surface or in the surrounding media.[206-208] Either the amplitude or phase signals can be used to extract thermal properties of thin films and interfaces.[209-211] Modulation of optical heating lasers are typically limited to several megahertz, but have also been pushed to hundreds of megahertz.

Transient thermoreflectance (TTR) is an optical pump-probe method that has become increasingly popular and powerful for thermal measurements. In brief, TTR shines high power pump laser pulses on a metal layer deposited on the sample. The electrons in the metal layer absorb the laser energy and rapidly convert it into lattice thermal energy through electron-phonon coupling. The thermal energy is transferred from the heated metal layer to the attached sample. The time history of the metal surface temperature is tracked by a delayed low-intensity probe laser through detecting the metal reflectance change. Using heat equations to fit the signal decay curve, the thermal conductivity of the sample can be obtained. More details of the experiments can be found in literatures.[211-213] Such a pump-probe technique is well suited to measure thermal transport properties of bulk materials and layered structures,[211-214] and can be used to detect transport at very short time scales. Besides the reflectance signal, transmission signal can also be used for such experiments.[215, 216]

This pump-probe technique, which was originally largely used for electron transport and electron-phonon interaction studies, was used by Paddock and Easely[217, 218] for the first time to measure thermal conductivity of thin metal films in 1986. Capinski and Maris[219] later improved this method by implementing a single-mode optical fiber after the delay stage to overcome a series of difficulties from the stage motion, alignment problems, and divergence of the beams. A number of other modifications, such as the use of femtosecond laser, were made by Norris and coworkers[213] and Cahill and coworkers[211, 220-223] to improve the signal-to-noise ratio by about ten times and extended its application to measure thermal interface conductance. Schmidt[224] further modified the technique by using a coaxial geometry of the pump and the probe beams to reduce the noise on the data acquired from a lock-in amplifier. Malen and coworkers[225] recently demonstrated a full frequency domain thermoreflectance technique that is capable of measuring thermal conductivity without the need of a delay stage, capitalizing on previous developments in the field of photothermal spectroscopy.[210, 226, 227]

Raman microscopy and photoluminescence have also been used to measure thermal conductivity of nanomaterials, such as graphene[8, 228] and nanowires,[229] or measure temperature distribution in devices. The measurement of temperature using Raman and photoluminescence depends on detecting the spectrum shifts which are functions of temperature. By measuring the shift in wavelength, temperature rise of the probed region can be inferred. However, the accuracy of this method is not very high as seen in the large error bars usually reported in these measurements.

### I. 2.6 Thermal Conductivity Spectroscopy

Information of phonon properties, especially mean free path and their contribution to thermal transport, is of great importance for predicting and engineering thermal transport properties of nanomaterials. A few experiments mainly using the pump-probe apparatus has made the mean free path information accessible experimentally. The common mechanism of these experiments is the classical size effect of phonon transport imposed by the finite size of the heated regions. When phonons have mean free path larger than these characteristic lengths, they are ballistic and their contribution to the heat flux is negligible.[230, 231] The measured thermal conductivity is the contribution from phonons with mean free path smaller than then characteristic lengths, which are diffusive.

The experiment carried out by Koh and Cahill[214] shows that the thermal conductivity of semiconductor alloys depends on the modulation frequency imposed on the pump beam when the frequency is larger than 1MHz. The thermal penetration depth (*h*) depends on the modulation frequency (*f*) according to Eq. (3):

$$h = \sqrt{\kappa/\pi f c} \qquad (3)$$

Here, the penetration depth becomes a characteristic length to divide diffusive and ballistic phonon transport (Fig. 3a). As a result, the measured thermal conductivity is approximated to be the contribution from all the phonons with mean free path smaller than the penetration depth *h*. By lowering the frequency which leads to larger penetration depth, contribution of longer mean free path phonons is included in the measured thermal conductivity. The thermal conductivity increments thus correspond to the contribution from these long mean free path phonons. Such a frequency dependent thermal conductivity has been predicted both at the molecular level[232] and the macroscale.[233] However, a recent simulation based on Boltzmann equation was not able to find such frequency dependence of thermal conductivity.[130]

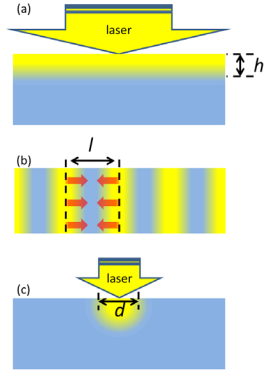

Figure 3 Characteristic lengths of different thermal conductivity spectroscopy methods: (a) thermal penetration depth, $h$, is the characteristic length in a modulation frequency varying transient thermoreflectance measurement; (b) grating period, $l$, is the characteristic length in a transient thermal grating method; (c) pump laser beam diameter, $d$, is the characteristic length in a spot size varying transient thermoreflectance measurement.

Later, Siemens et al.[234] used ultrafast infrared laser to heat up a nickel grating pattern deposited on a sapphire substrate. They used a coherent soft X-ray beam to probe the heat flow dynamics by monitoring the diffraction intensity as a function of time. The relaxation curve is then fitted using heat conduction models to evaluate thermal resistivity. Through changing the width of the nickel pattern, the size dependent thermal resistivity is measured. In this experiment, the grating periodicity $l$ works as a characteristic length (Fig. 3b). Johnson[207] used another pump-probe method called transient thermal grating to measure the in-plane thermal conductivity of thin films. After passing the pump laser through a pair of achromatic doublets, the two pump beams, which was previously split using a phase mask, cross at the surface of the graphene sample, forming a sinusoidal grating pattern with a period of $l$ due to interference, with $l$ being determined by Eq. (4):[206]

$$l = \lambda / \left(2\sin\frac{\theta}{2}\right) \qquad (4)$$

where $\lambda$ is the wavelength of the pump laser and $\theta$ is the crossing angle. The peaks of the interference pattern induce locally heated regions and result in transient thermal grating (Fig. 3b). The material's response to this transient grating is monitored by a photo detector into which the diffracted probe beam is directed. By probing the dynamics of the intensity of the diffracted laser, thermal decay is monitored, and the thermal conductivity is extracted through fitting the decay using a heat conduction model. The periodicity, $l$, imposes a characteristic length to divide ballistic and diffusive phonon transport. By changing the periodicity and measuring the effective thermal conductivity, the contribution of phonons with different mean free paths can be studied. Reasonable agreement between the $l$ dependent thermal conductivity from such experiments and that from calculations has been observed.[207]

Minnich et. al[85] also used a TTR setup but used the pump laser diameter ($d$) as the characteristic length to probe the phonon mean free path information (Fig. 3c). The dependence of thermal conductivity on the laser diameter was observed. The difference of the thermal conductivities between two neighboring ($d_1$->$d_2$) diameters is attributed to the phonons with mean free path between $d_1$ and $d_2$. They were able to measure the cumulative thermal conductivity of silicon with respect to the phonon mean free path and found good agreement with that from the aforementioned first-principles approach.[45]

One challenge of the thermal conductivity spectroscopy is how to further minimize the characteristic lengths used in the experiments. This is necessary for probing the mean free path dependent thermal conductivity of low thermal conductivity materials. The thermal penetration depth, which is inversely proportional to the modulation frequency, can be minimized using higher modulation frequencies. However, the signal to noise ratio decreases significantly when the modulation frequency becomes too high, impairing the measurement accuracy. The thermal grating period can be lowered using small wavelength lasers and large crossing angles according to Eq. (4). However, the choice of wavelength is usually limited by the required absorption rate of the material, and the crossing angle is limited by the practical features of the achromatic doublets. On the other hand, the practical pump spot size can only be lowered to a few hundreds of nanometers. It is possible to use metal layers patterned to much smaller sizes using lithography instead of a whole metal layer as the photo-thermal transducer. To make this method effective, only the patterned metal film should absorb laser energy. However, many materials such as $Bi_2Te_3$ also has significant absorption rate in the wavelength range in which the optical resonance wavelength of metals lies. Coating the sample with a low absorption rate material may be a possible solution to this problem, but this will complicate the sample preparation and data analysis.

## II. Novel Thermal Transport Phenomena and Properties

The above-discussed computational and experimental techniques have enabled the observation, understanding and prediction of many novel thermal transport phenomena especially at the nanoscale, stimulated the discovery of nanomaterials with novel thermal transport properties. In this section, we will give examples of novel heat thermal transport phenomena in both hard and soft materials, across interfaces, and also briefly comment on photon transport at nanoscale.

### II. 1 Hard Materials

#### II. 1.1 Low Dimension Materials

Silicon nanowire, a quasi-one-dimensional material, displays a thermal conductivity which is significantly lower than bulk silicon.[30, 235] Compared to the three-dimensional bulk silicon, the small cross-sectional area of nanowires imposes classical size effect to phonon transport. The classical size effect of phonon transport refers to a phenomenon in which phonon scattering at the structure boundaries or interfaces are enhanced when the structure sizes are smaller or comparable to the intrinsic phonon mean free paths. Such an effect reduces the effective mean free path of phonons. Through a number of theoretical and computational studies,[81, 236-238] it is now widely accepted that the enhanced boundary phonon scattering at the nanowire surfaces, which significantly lower the mean free paths of phonons, is the main reason that lead to the lowered thermal conductivity. Hochbaum et al.[31] was able to further reduce the nanowire thermal conductivity by roughening the surface which enhanced the boundary scattering. The low thermal conductivity they achieved was about 100-fold lower than that of bulk silicon and cannot be explained by the classical size effects models. A recent study by the same group showed that considering phonon wavelength is important in explaining the observed thermal conductivity in these silicon nanowire, and the thermal conductivity reduction is correlated well with the power spectra of surface roughness.[32] Such a reduced thermal conductivity has inspired the idea of using silicon nanowire for thermoelectrics application.[31, 239]

One unexpected phenomenon – the linear temperature dependence of the thermal conductivity in a 22nm silicon nanowire,[30] however, has not been explained satisfactorily. Murphy and Moore[240] argued that such a phenomenon can be understood through a model that combines incoherent surface scattering for short-wavelength phonons with nearly ballistic long-wavelength phonons. This model was able to fit the linear thermal conductivity for small diameter nanowires but lacked accuracy for larger diameter nanowires especially at high temperatures.[241]

Thermal conductivity reduction has also been found in thin films. Goodson and coworkers[161, 242-245] used different measurement techniques, such as 3$\omega$ and optical thermoreflectance methods, to find that both the in-plane and out-of-plane thermal conductivities of thin films are much lower than the bulk counterpart. Similar trend has also been observed by Chen and coworkers.[246] The effect of boundary scattering on thermal conductivity reduction in thin film structures has also been studied by theoretical works.[247-249] All data on reduced thermal conductivity in thin films can be well explained based on the Casimir classical size effects picture,[250] especially the Fuchs-Sondheimer models.[251, 252]

Superlattices made of periodically alternating thin films have shown significantly reduced thermal conductivity compared to their corresponding bulk alloys,[24, 25, 27, 253-255] and such thermal conductivity reduction has led to reports in enhanced ZT in $Bi_2Te_3/Sb_2Te_3$ and PbTe/PbSe superlattices.[256, 257] The reduction of thermal conductivity in superlattices compared to the bulk alloy has been theoretically studied. Two major pictures exist, one based on the coherent[258] and the other on incoherent[247, 259] phonon transport. In the coherent transport picture, the main reason for thermal conductivity reduction is the decrease in the phonon group velocity, while in the incoherent picture the reason is phonon scattering at the interfaces. However, none of these two effects can fully explain the experimental observations until they are combined.[255] It has been found that the interface roughness, which significantly enhance phonon scattering compared to a planar interfaces, is the main reason for large reduction achieved in superlattice thermal conductivity.[247, 260] A more recent experiment shows that coherent phonon transport in superlattices – heat carried by extended Bloch waves through the whole thickness of the superlattices – contributes to a significant fraction of the measured thermal conductivity since interface roughness cannot effectively scatter long wavelength phonons.[261] If the coherence of these phonons can be destroyed, further reduction in the thermal conductivity is possible. Chen argued that effective thermal conductivity of superlattices can be reduced to be lower than the value predicted by the minimum thermal conductivity theory.[29] Chiritescu et al.[262] experimentally observed this breakdown of the amorphous limit in a disordered layer structure of $WSe_2$ which has a thermal conductivity six times lower than the amorphous limit. More surprisingly, after the sample was subject to ion-irradiation, which is generally used to introduce disorder to lower thermal conductivity, the sample exhibits a five-fold increase in thermal conductivity. The authors speculated that the low thermal conductivity of disordered layers was produced by random stacking of well-crystallized $WSe_2$ sheets which increase phonon scattering. Molecular dynamics simulations[262] using a model potential showed that the low thermal conductivity is related to the very week inter-layer van der Waals interactions between $WSe_2$ sheets. These pictures, however, cannot explain the increased thermal conductivity after ion-irradiation.

The above discussed thermal conductivity reductions in nanowires, thin films, and superlattices are caused by the interface scattering, which are more or less captured by the Casimir classical size effect picture. As the size further shrinks, a nanowire become a molecular chain, and a thin film into a molecular sheet, transport physics change. The famous Fermi, Pasta and Ulam little discovery suggests that one-dimensional atomic chain could have infinite thermal conductivity.[6, 263] The molecular dynamics simulation by Berber et al. in 2000[2] calculated a very high thermal conductivity (~6600W/mK at room temperature) of single-walled carbon nanotube – another quasi-one-dimensional nano material. Such high thermal conductivity exceeds the bulk counterpart of carbon structures – diamond, showing a trend opposite to that would have been predicted by the classical size effect. This work inspired significant interest in the thermal transport of low dimensional materials. In the following year, thermal measurement was successfully performed by Kim et al.[163] on a standalone multi-walled carbon nanotube using a microfabricated platform, and they confirmed a very high thermal conductivity of more than 3000W/mK at room temperature. In 2005, Pop and coworkers[3] did similar experiment on a single-walled carbon nanotube and also observed exceptional thermal conductivity of nearly 3500W/mK at room temperature. These experimental findings also motivated the exploration on the mechanism of the ultrahigh thermal conductivity in such one-dimensional materials. Li et al.[264] found that the high thermal conductivity is due to anomalous heat diffusion in a carbon nanotube which can lead to diverging thermal conductivity. Such a diverging thermal conductivity has also been found in silicon nanowires using a molecular dynamics simulation.[265] Shiomi and Maruyama[266] used molecular dynamics to find that thermal conductivity diverges with respect to the simulated carbon nanotube length even when it is over 1μm, suggesting that phonons have very long mean free paths. It is later found by Lindsay and coworkers[5] from lattice dynamics calculations that the Umklapp scatterings, which lead to thermal resistance, can hardly happen due to the one-dimensional feature of the phonon dispersion relation which makes the scattering selection rules difficult to satisfy. This explains the fundamental difference between the one dimensional phonon transport and classical size effect. In the case of diverging thermal conductivity, one cannot use perturbation method and Boltzmann equation to predict heat transfer.

Graphene is a two-dimensional material consists of a monoatomic carbon layer.[7] The thermal conductivity of a suspended single layer graphene measured by Balandin et al.[8] using confocal Raman spectroscopy was found to be as high as 5800W/mK at room temperature, even higher than carbon nanotube. High thermal conductivity was also later reported on both suspended[228] and supported graphene.[168] It has been observed that the supported graphene has thermal conductivity a few folds lower than suspended graphene, which is due the phonon leakage and scattering resulted in by the substrate.[168] Well before the discovery of the high graphene thermal conductivity, Klemens[267] predicted a high thermal conductivity of 1900W/mK of such two-dimensional carbon network. Balandin and coworkers[268] used Klemens' model with the Grüneisen parameters calculated from DFT and found good agreement between their calculated data and the experimental data. In Klemens' model, in-plane modes including longitudinal acoustic (LA) modes and transvers acoustic (TA) modes were assumed to dominate while the out-of-plane acoustic modes (ZA) have negligible contribution due to their large Grüneisen parameters and small group velocity near Brillouin zone center. However, Lindsay et al.[74] used Boltzmann equation with an exact numerical solution to find out that the ZA modes in graphene dominate the thermal transport, which is related to the unusually strong Normal scattering and their large density of state near Brillouin zone center.

The ultrahigh thermal conductivity graphene start to find applications in thermal management of electronics.[269] It is found that a graphene heat spreader was able to lower the hot spot temperature of a GaN transistor by 20 degrees, which can potentially elongate the device lifetime by an order of magnitude.[270] Such applications involve thermal transport in the cross-plane direction of graphene and across its interfaces with other materials. Dames and coworkers[271] used the $3\omega$ method to measure thermal contact resistance between silicon dioxide and single and few layer graphene. They found low thermal resistance at these interfaces ($5.6\times10^{-9} - 1.2\times10^{-8}$ m$^2$K/W) which shows no clear dependence on the graphene thickness. Koh et al.[272] used TTR experiment to find out that graphene between two solid surface presents larger thermal resistance. Such resistance increases as the graphene thickness increases from one layer to three layers but shows no strong thickness dependence after that.

### II. 1.2 Bulk Materials

Bulk materials, through nanoengineering, have also displayed novel thermal transport properties, especially low thermal conductivity. The motivation of engineering low thermal conductivity materials have largely stemmed from thermoelectrics research. By introducing nano-sized grains into a Bi-Sb-Te bulk alloy, its peak figure of merit, ZT, has been improved by 40%.[34] The most importance contribute of nanograins to this improvement is that the long mean free path phonons are scattered by the grain boundaries, leading to a significant reduction in thermal conductivity.[34] However, what is the right nano-grain size that can most effectively scatter phonons? Phonon mean free paths vary from one material to another. Different phonon modes can have very different mean free paths and contributions to the total thermal conductivity. Motivated by these questions, computational work, including molecular dynamics simulations and first-principles calculations, has been done to calculate the mean free path of phonons and study their contributions to thermal transport in a range of semiconductors, including silicon,[45] GaAs,[58] half-heusler,[59] PbSe[69] and PbTe[67, 69] (Fig. 4). It was surprising to find out from molecular dynamics results[91] that the phonon mean free path of silicon can spanned over six orders of magnitude and over half of the thermal conductivity is contributed by phonons with mean free path longer than 400nm even at room temperature. More accurate first-principles calculations[45] predicted this half-way mark to be 1µm. These findings explained the high temperature size effect found in silicon thermal conductivity around the 70s.[273] Along these advances, experimental techniques such as thermal conductivity spectroscopy and inelastic neutron scattering can provide experimental validation to simulations.[85]

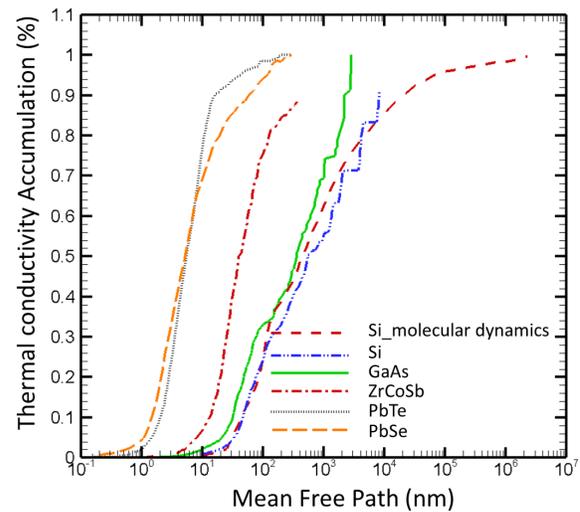

Figure 4 Thermal conductivity accumulation with respect to phonon mean free path in different materials.(Si_molecular dynamics[91]; Si[45]; GaAs[58]; ZrCoSb[59]; PbTe[69]; PbSe[69])

For low thermal conductivity materials, however, the phonon mean free paths were found to be much smaller. For example, about 90% of the PbTe thermal conductivity is contributed by phonons with mean free path smaller than 10nm, making reduction of thermal conductivity more difficult than silicon. However, nanoparticle sizes below 10nm have been achieved by solid solution alloying of PbTe−PbS and thermal treatment which induces a spontaneous nucleation and growth of PbS nanocrystals. This has led to a reduction of PbTe thermal conductivity by 60% at 400 and 500K.[274]

### II. 2 Soft Materials

### II. 2.1 Polymers

Amorphous polymers, which are generally regarded as thermal insulators, have very low thermal conductivities within a relatively narrow range (0.1-0.5W/mK).[9] These polymers, however, are being used in thermal applications such as thermal interface materials in microelectronics where high thermal conductivity is needed. The general understanding of thermal transport in amorphous polymer is limited. The model proposed by Allen and Feldman was able to explain thermal conductivity behavior in disordered inorganic materials such as $\alpha$-Si.[275] In amorphous silicon, all bonds are the same, meaning that the environment of each atom is more or less equivalent and isotropic. However, in amorphous polymers, atoms experiences highly anisotropic forces, including bond forces along the polymer chain and nonbond forces (i.e., van der Waals and Coulombic) across different chains. Heat transfer can be very different if they are along the chain backbone which is connected by strong covalently bonds versus if they are transferred across the chains with the aid of much weaker van der Waals or Coulombic forces. Theoretical work considering these factors is currently missing and is important toward the full understanding of the thermal transport in polymers.

Although amorphous polymer has low thermal conductivity, recent experiment shows that polymers can be exceptional thermal conductors if the polymer chains are straight and aligned in the crystalline fiber form.[11, 276] The ultra-draw polyethylene nanofibers were found to have thermal conductivity as high as ~100W/mK, about three hundred times higher than that of the amorphous counterpart. The mechanism of such remarkable thermal conductivity was explored through molecular dynamics simulations, and it is found that the root cause is the very high or even divergence thermal conductivity of polyethylene single chains.[10] Similar to CNTs, phonons in single polymer chains can travel significant distances without attenuating due to the one dimensionality. The divergent thermal conductivity in polyethylene single chains was found to be contributed by the non-attenuating cross-correlation of the energies of different phonon modes.[92] Such non-attenuating phonons were first proposed as the Fermi-Pasta-Ulam problem[6] for a one-dimensional chain. It was predicted in a one-dimensional nonlinear chain, energy can cycle through different phonon modes, resulting in non-attenuating energy transport.

It is also found that the thermal conductivity of polymer chains exhibit a completely different trend from silicon when they transfer from one-dimensional chains to a three-dimensional crystal. In silicon, thermal conductivity decreases as the structure changes from a three-dimensional bulk to a one-dimensional wire due to the aforementioned classical size effect. However, in polymer, thermal conductivity decreases when it is expanded from a single chain to a crystal bundle. Such a trend was found to be caused by the inter-chain interactions, which work as scattering sources for phonon transport in each individual chains.[277]

However, not all polymer chains have high thermal conductivity. It is found that the thermal conductivity of polymer chains is a strong function of their morphology. For example, PDMS single chains have thermal conductivity around 7W/mK,[278] significantly lower than that of polyethylene. The segmental rotation due to the weak dihedral angle of the PMDS chain backbone imposes strong disorder scattering to phonons and thus lower the thermal conductivity. It is also recently observed from molecular dynamics simulations that such segmental rotation can cause the thermal conductivity of a polyethylene crystalline fiber to drop by a factor of ten suddenly around 400K.[279] It is found that thermal expansion which creates more space for each individual chain in the crystalline structure leads to easier chain segmental rotation. Such an effect due to segmental rotation has also been found in polymer chains with different monomers.[280]

**II. 2.2 Nanofluids**

Nanofluids, fluids with high thermal conductivity nanoparticle suspensions, are attractive for heat transfer research due to their potential applications such as high performance coolant and refrigerant. In the 1990s, experiment has shown that thermal conductivity of liquid can be enhanced significantly by adding nano-sized particles into the liquid to form suspensions.[12, 13] The thermal conductivity enhancement is much higher than that predicted from traditional effective medium theories assuming well-dispersed isolated particles in the fluids.[14-19] It is, however, argued that the enhancement is within the limit of the Maxwell theory[141] when considering a continuous phase of particles which form percolating structures.[281] Other reports could not reproduce the enhanced thermal conductivity. Past controversies on the enhancement mechanism and validity of experimental data, have been discussed elsewhere.[282] We will discuss below few recent observations that shed light on the heat conduction mechanisms together with some phenomena..

Nanoparticles suspended in fluids are not isolated but form internal structures.[283-285] It has been observed that when alumina nanoparticles are dispersed in hexadecane, particles form clusters. When the nanofluid is frozen, the clusters form long range network as the particles are pushed to the grain boundaries when hexadecane form crystal domains in the frozen state (see TEM image in Fig. 5 a & b).[286] Such a morphology change leads to an increase in the thermal conductivity. On the contrary, a slight decrease in the thermal conductivity is observed in hog fat based nanofluids. It is found that large scale percolation network is not formed in hog fat based nanofluids (Fig. 5 c & d). These observations support the conclusion that heat conduction along solid structures, which could form complicated clusters and/or percolation network, is the mechanism of the anomalous enhancement of thermal conductivity in nanofluids, rather than Brownian motion as suggested in other literatures.

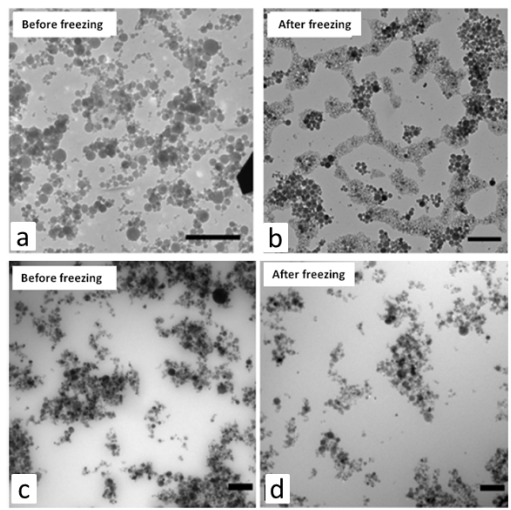

Figure 5 TEM images of alumina nanofluids using hog fat (a and b) and hexadecane (c and d) before and after freezing.[286]

As heat conduction along the solid clusters holds the key for thermal conductivity enhancement in nanofluids, graphite flakes and carbon nanotubes with high thermal conductivity along the basal plane are ideal candidates for enhancing heat conduction. Graphite flakes have also been suspended in ethylene glycol and other liquids to form nanofluids using ultrasonic processes. Such compositions have led to increases in the thermal conductivity. Moreover, it is observed that there are kinks in the thermal conductivity enhancement curves as a function of volume fraction. These kinks are believed to be related to the particle percolation (Fig. 6).[287] Although electrical percolation is well known, distinct thermal percolation had not been observed in other materials. The slopes in thermal conductivity increase before the kinks are larger than those after the kinks, which is contrary to electrical percolation for which electrical conductivity increases in power law above the percolation threshold. Based on optical microscopy and AC impedance spectroscopy experiments,[287] it is concluded that at low volume fractions, the graphite flakes in isolated clusters are more closely bonded due to the effect of surface energy minimization. This is favorable for the intra-cluster thermal transport. When percolation happens, the clusters merge into larger networks and the driving force to minimize the surface energy becomes smaller. As a result, the graphite flakes bond more weakly, leading to larger interfacial thermal resistance.

As the evidences show that thermal transport along the percolated solid state network is the key to the thermal conductivity enhancement, we can say that high thermal conductivity additives are preferable but the additives should also be selected so that they form stable network in nanofluids. Particles with high aspect ratio are also preferred since they minimize the contact resistance within the same distance in the network.

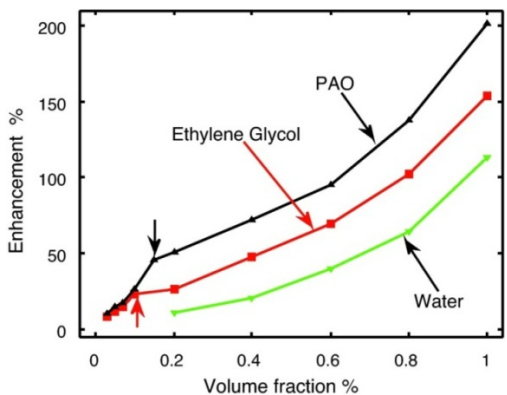

Figure 6 Thermal conductivity enhancements in graphite nanofluids as a function of volume fraction.[282]

An important aspect to be emphasized is that the observed enhancement strongly depends on how the nanofluid is prepared which greatly influences the cluster structures in the suspension.[283, 285] As a result, more thorough understanding of the structure of particulates in liquids, including how thermal contact resistances varies with structural change, is critical to precisely control and fully understand the thermophysical properties of nanofluids. In fact, the study of the structures of nanoparticles in a liquid falls into the realm of soft materials, a topic that has attracted tremendous interest in condensed matter physics. Further research relating the structure of soft materials and their transport properties is expected to be fruitful.

Based on the freezing experiment, a reversible temperature regulation of suspension's electrical conductivity and thermal conductivity has been achieved.[288] The achieved contrast ratio of thermal conductivity between the liquid state and the solid state is three times, and that of electrical conductivity is as large as four orders of magnitude. Such material properties can enable potential applications such as building temperature regulation, thermal storage, sensors and thermal fuse.

## II. 3 Interfaces

As the dimensions of structures reaching nanometer, thermal boundary resistance becomes increasingly important due to the large density of interfaces. Theoretical prediction of thermal transport across interfaces has been persistently difficult. There are empirical models, such as the acoustic-mismatch model (AMM)[289] and the diffusive mismatch model (DMM).[290] These models, however, lack accuracy and can sometimes be off by orders of magnitude.[35, 290] One of the most important features that cannot be included properly in the models is the microscale structure and interaction at the interfaces, which are recently shown to strongly influence thermal conductance of both solid-solid interfaces and solid-soft material interfaces.

### II. 3.1 Solid-Solid Interfaces

A series of recent experiments have shown that interfacial morphology and surface chemistry can have great impact on interfacial thermal transport. Collins et al.[291] found that the interfacial conductance of a oxygenated diamond-Al interface can be as large as four times greater than that of a hydrogenated counterpart. Hopkins et al. also found graphene with oxidization groups results in higher thermal conductance at its interface with Al compared to pristine graphene.[292] A possible reason is that the some chemical groups at the sample surfaces facilitate the formation of stronger chemical bonds between the metal and the sample, and thus enhance thermal transport. A theoretical model, which is based on the acoustic mismatch model but accounts for the adhesion energy of the interface, has been developed by Prasher.[293] This model relates the interface bond strength with interfacial thermal transport and is able to qualitatively explain the bond strength dependent interfacial resistance found in some experiments. This model, however, assumed perfect atomic arrangement at the interfaces.

Besides surface chemistry, interface microstructure also plays a role in thermal transport. Hopkins et al.[294] prepared different Al-Si interface with the silicon surfaces subjected to different surface treatments, which effectively changes the surface roughness to different extend prior to the aluminum deposition. It was observed that the thermal conductance decreases with the increase in surface roughness. It was also reported that if the native oxide layer on the silicon surface is stripped off, the thermal conductance increases. This seems to contradict with the trend found in references,[291, 292] but was supported by another experiment[295] which found that the thermal conductance of Al-SiC interface is larger than that of Al-SiO$_x$-SiC interface. It is possible that the oxide layer presents large thermal resistance. The full understanding of these experimental observations, however, requires more detailed studies.

Hsieh et. al[295] also found that the thermal conductance at the Al-SiO$_x$-SiC increases by 4-5 times when a pressure of 10GPa is applied, while only very slight change is observed for the Al-SiC interface subject to this pressure. The authors speculated that the Al-SiO$_x$-SiC interface is weakly bonded and thus is more susceptible to external pressure. Similar phenomena have been observed by Luo and Lloyd[112, 115] in their molecular dynamics simulations, which show that for an interface bound by strong covalent bonds, the external pressure has little influence on the interfacial thermal transport,[112] but for a weak interface bound by van der Waals interactions, interfacial thermal conductance is a strong function of pressure.[115]

It is worth noting that most measured interfaces using pump-probe and 3ω methods are between metal and dielectrics due to the limitations of the experiments. These interfaces are usually subject to strains and defects, and their interfacial thermal conductance is usually on the order of 10-100MW/m$^2$K. For cleaner interfaces such as those in superlattices, thermal conductivities are higher.[260]

**II. 3.2 Solid-Soft Material Interfaces**

Thermal transport across solid-soft materials interfaces is another area where engineering and theoretical interests reside. Characterizing the relationship between the interfacial interaction and conductance has been performed by several groups. Thermal conductance between solid and liquid with different surface termination molecules changing the hydrophobicity has been measured by Ge and Cahill.[296] It is found that hydrophilic interfaces have thermal conductance (100-180MW/m$^2$K) 2-3 times larger than that of hydrophobic interfaces (50-60MW/m$^2$K). Such a finding is confirmed by Shenogina et al.[297] using molecular dynamics simulations. The hydrophilic surface enhances the interfacial interaction between solid and liquid and thus facilitates thermal transport. However, another experiment by Ge et. al[298] measuring the interfacial thermal conductance between metallic particles and solvent shows no strong dependence of the conductance on the surface functional groups. The possible reason is that these functional groups used to stabilize the nanoparticle suspension in the solvent all have similar hydrophilic head groups, and it is the interaction between these head groups and solvent molecules that dominates thermal resistance. Schmidt et al.[299] found that the interfacial thermal conductance between CTAB coated gold nanorods and solvent is a function of the surfactant concentration.

Solid-polymer interface is another kind of solid-soft material interface which has also been studied. Losego et al.[300] used TTR to measure interfacial effects in the thermal conductance of spun-cast polymer thin films and polymer brushes grafted on substrate. The grafted interface, which should has stronger interfacial bonds and is expected to have polymer chains partially aligned perpendicular to the substrate, only has a modestly higher (13%) thermal conductance than the spun-cast layers. A recent molecular dynamics simulation,[115] however, showed that the formation of covalent bonds can enhance interfacial thermal conductance by almost one order of magnitude compared to van der Waals interfaces. Such an effect has been confirmed by another experiment performed by Losego et al.[301] who studied the relation between the chemical bonding and interfacial thermal conductance. By systematically increasing the density of covalent bonds at the interface, thermal conductance was found to increase accordingly. These findings agrees with Luo's previous molecular dynamics simulations.[112]

**II. 4 Near-Field Radiation**

So far, our discussion has focused on heat conduction. Thermal radiation at nanoscale also has shown very distinct departure from the Planck's law of blackbody radiation.[302] In the far field, radiation heat transfer between macroscopic bodies is limited to below that given by the Plank blackbody radiation law for the spectral-based flux or the Stefan-Boltzmann law for the total flux. When two objects are brought close to each other, radiation can exceed the blackbody limit because of the tunneling of evanescent waves.[20, 21] There are two types of evanescent waves.[36, 303, 304] One decays exponentially from one-side of the interface but is a propagating wave on the other side of the interface, and the other decays exponentially on both sides of the interface. The former happens when electromagnetic waves inside a material with a high and positive dielectric constant propagate cross the interface with a material of a lower but positive dielectric constant at an angle of incidence larger than the critical angle as determined by the Snell law, creating an exponentially decaying electromagnetic field, i.e., an evanescent wave, on the other side. The latter is called a surface wave which happens when the dielectric constants on both sides of an interface are opposite signs, and the absolute value of the negative dielectric constant is larger than that of the positive dielectric constant. A negative dielectric constant occurs when the frequency of photons falls in between the transverse and longitudinal optical phonon frequencies for dielectrics or below the plasma frequency for metals, when there are resonances between material waves (e.g., optical phonons, electrons, etc.) and the electromagnetic waves. Examples of surface waves are surface plasmon-polaritons or surface phonon-polaritons, which are resonance modes of the electrons and electromagnetic waves, and the optical phonons and electromagnetic waves, respectively. Evanescent waves normally do not carry a net energy flux and decays rapidly away from the interface in a distance of the order of the wavelength, and hence do not contribute to radiation heat transfer between objects separated at macroscopic distances. However, when the two surfaces are brought close by, evanescent waves can reach the other medium and enhance heat transfer.

Measurement of near-field radiation is challenging because (1) thermal radiation heat transfer rate is small and (2) near-field effects start to be important only when the separation between the two surfaces are comparable to the dominant wavelength. Measuring small heat transfer at small separations of two surfaces is difficult. Tien and co-workers[20, 305] reported near–field radiative heat transfer at cryogenic temperatures between two copper disks. Due to the low emittance of metallic surfaces, the maximum heat transfer observed is ≈ 1/29 of the Planck theory prediction. Hargreaves[306] carried out near-field radiation heat transfer measurements between two chromium surfaces with a diameter of 4.0 cm$^2$ with gaps down to 1 μm at room temperature, demonstrating an enhancement from ≈ 1.75 Wm$^{-2}$K$^{-1}$ in the far–field to ≈ 2.95 Wm$^{-2}$K$^{-1}$ at a 1 μm gap, which is ~50 % of blackbody radiation. More recently, Kralik et al.[307] carried out near-field measurements between tungsten surfaces, each 35 mm in diameter, with a separation distance between ~ 1 μm to 1 mm. The hot-side surface was maintained between 10-100 K and cold side maintained at 5 K. They reported a three orders of magnitude increase in the emittance between the two surfaces. To push to smaller separation between surfaces, Xu et al.[308] took the strategy of reducing the surface area of two metallic surfaces using a scanning tunneling microscope(STM) configuration. One side of the surface is a thin-film thermocouple, and the other side of the surface is an indium tip flattened by pressing it against a glass flat. Their experiment, however, was not conclusive due to the significantly reduced heat transfer with smaller surface area and the low sensitivity of the thermocouple heat flux sensor. Kittel et al.[309] used an improved STM configuration to measure heat transfer of a gold tip with a gold substrate and with a GaN substrate for tip-substrate separation in the range of 1-100 nm. They observed saturation of heat transfer when the tip-substrate separation is less than 10 nm.

To study near-field heat transfer between non-metallic surfaces, especially between two surfaces supporting surface phonon-polaritons, Hu et al.[310] used two quartz surfaces with a diameter of 1.27 cm separated by polystyrene beads with a nominal diameter of 1 micron. They reported the near-field heat transfer between the quartz surfaces exceeding that of the blackbody radiation. However, their experiment cannot be pushed for smaller separations, at which additional thermal shorts, which could arise from the physical contact due to diameter variations in the polystyrene beads or other dust particles. Recently, Ottens et al.[311] reported measurements of near-field radiation between two large sapphire surfaces, 5 cm x 5 cm in surface area, for distance between 2-100 μm near room temperature. They maintained the surface separation and parallelism using small capacitive sensors deposited on the surfaces. Their experiment observed increasing heat transfer with decreasing separations between two surfaces starting from 10 μm, and heat transfer coefficient exceeding that between blackbodies.

The Chen group at MIT developed a sphere-plate experimental configuration to investigate near-field heat transfer.[23, 312] This experimental configuration overcomes the difficulties of maintaining parallelism in the two parallel plates, and unknown tip geometry in the STM experiment. Their experiment took advantage of the proven extreme thermal sensitivity of bi-layer cantilevers used for atomic force microscopes,[193, 313, 314] which compensates for a smaller area involved in radiative transfer due to the curvature of the sphere compared to the parallel plate configuration. They reported experimental results between a glass sphere and different substrates including glass, silicon, and gold, for sphere and plate separation down to 30 nm. Such experiments were extended to smaller separations and between metallic surfaces.[315, 316] Rousseau and co-workers used a slightly modified configuration and reported near-field heat transfer between a glass sphere and a glass substrates.[317] A subtle but significant difference between the two experiments is that in the MIT experiment, the cantilever is heated while the surrounding is maintained at ambient temperature, while the Rousseau et al.'s experiment heated the substrate. In the former case, only near-field heat transfer is measured, since the far-field contribution is included in the initial bending of the cantilever. In the latter case, the influence of the heated substrate to the sphere, and cantilever is modeled and used as a fitting parameter. Theoretical simulation by Otey and Fan[318] shows that the far-field component creates significant ambiguity to the data interpretation.

Future research could further push the objects even closer which might achieve even higher heat transfer coefficients. However, when the two objects are very close, current electromagnetic theories are no longer valid. In such a region, heat conduction mediated by atomic interactions at the interface becomes important. The transition from heat radiation to conduction is a fundamental problem that is of great scientific interest.

## III. Summary

In this review, we have discussed recent progress in computational and experimental techniques in nanoscale thermal transport. Lattice dynamics based on first-principles calculations has shown great prediction power in thermal conductivity calculations for perfect crystals. Molecular dynamics simulations using empirical potential models will continue to be an important tool for studying complicated structures. Accurate empirical potential models hold the key to the accurate prediction of thermal transport properties. Reactive potentials are promising in offering more realistic description of interface and surface structures which are found to be critical in interfacial thermal transport. Multiscale simulations utilizing results from atomistic simulations can model thermal transport of systems at larger scales to bridge the gap between atomistic simulations and device level properties. However, interfacial phonon transmission is still difficult to obtain in the current atomistic simulation framework. Experimental techniques have also advanced significantly. Microfabricated platforms have been very successful in measuring thermal conductivity of nanostructures. Various optical pump-probe methods are being widely used in measuring interfacial thermal conductance as well as thermal conductivity of materials. Thermal conductivity spectroscopies based on pump-probe methods have been developed, enabling the measurements of phonon mean free path dependent thermal conductivities utilizing classical size effect. These techniques have been successfully demonstrated in materials with high thermal conductivities. The challenge, however, exists in expanding the applicability to measure materials with low thermal conductivity and short phonon mean free paths.

Novel thermal transport phenomena and properties observed in recent years are also reviewed. Low dimensional materials have exhibited interesting thermal transport properties including exceptionally high thermal conductivity in CNT, graphene, single polymer chains and significantly reduced thermal conductivity in thin films and nanowires. Thermal conductivity reduction has also been achieved by methods like nanocompositing, surface roughening and superlattices. Such reduction has enabled engineering applications such high performance thermoelectrics. Nanofluids have much higher thermal conductivities than those predicted by traditional theories. Recent experiments have led to the understanding of the importance of nanoparticle clustering in the thermal transport of nanofluids. By manipulating percolation, thermal conductivity enhancement has been achieved in a switchable and reversible manner, leading to a variety of potential applications. Radiative thermal transport at the near-field was found to be orders of magnitude larger than Planck's law of blackbody radiation.

Despite the significant advances in the field, there are still challenges existing in understanding the mechanism of nanoscale thermal transport phenomena. Current theoretical, computational and experimental tools needs to be further improved and new ones need to be developed to tackle these challenging problems. We hope that this review can call the attention of researchers from the physical chemistry and chemical physics communities into this exciting area of research.

## Acknowledgements

The authors thank Professor M. Zabarjadi for sharing part of the data shown in Fig. 1. T. L. thanks the support from Notre Dame startup fund. The computations are supported in part by NSF through XSEDE resources provided by SDSC Trestles under Grant No. TG-CTS100078. G.C. acknowledges partial supports from Solid-State Solar-Thermal Energy Conversion (S$^3$TEC) Center – a DOE Energy Frontier Research Center – on first-principles simulation (DE-SC0001299), DOE BES on near-field heat transfer (DE-FG02-02ER45977), and AFOSR on soft matters (FA9550-11-1-0174).

## Notes and references

$^a$ *Department of Aerospace and Mechanical Engineering, University of Notre Dame, Notre Dame, IN 46556, U.S.A. Fax: 5746318341; Tel: 5746319683; E-mail: tluo@nd.edu*
$^b$ *Department of Mechanical Engineering, Massachusetts Institute of Technology, Cambridge, MA 02139, U.S.A. Fax: 6173245519; Tel: 6172530006; E-mail: gchen2@mit.edu*
\* Corresponding Author

1. Chen, G. Nanoscale energy transport and conversion: a parallel treatment of electrons, molecules, phonons, and photons (Oxford University Press, 2005).
2. Berber, S., Kwon, Y.-K. & Tománek, D. Unusually High Thermal Conductivity of Carbon Nanotubes. *Physical Review Letters* **84**, 4613-4616 (2000).
3. Pop, E., Mann, D., Wang, Q., Goodson, K. & Dai, H. Thermal Conductance of an Individual Single-Wall Carbon Nanotube above Room Temperature. *Nano Lett* **6**, 96-100 (2005).
4. Shiomi, J. & Maruyama, S. Molecular dynamics of diffusive-ballistic heat conduction in single-walled carbon nanotubes. *Japanese Journal of Applied Physics* **47**, 2005-2009 (2008).
5. Lindsay, L., Broido, D.A. & Mingo, N. Lattice thermal conductivity of single-walled carbon nanotubes: Beyond the relaxation time approximation and phonon-phonon scattering selection rules. *Physical Review B* **80**, 125407 (2009).
6. Fermi, E.P., J.; Ulam, S. (Los Alamos Lab, 1955).
7. Novoselov, K.S. et al. Electric Field Effect in Atomically Thin Carbon Films. *Science* **306**, 666-669 (2004).
8. Balandin, A.A. et al. Superior thermal conductivity of single-layer graphene. *Nano Lett* **8**, 902-907 (2008).
9. Sperling, L.H. (Wiley-Interscience, 2006).
10. Henry, A. & Chen, G. High Thermal Conductivity of Single Polyethylene Chains Using Molecular Dynamics Simulations. *Physical Review Letters* **101** (2008).
11. Shen, S., Henry, A., Tong, J., Zheng, R. & Chen, G. Polyethylene nanofibres with very high thermal conductivities. *Nat Nanotechnol* **5**, 251-5 (2010).
12. Masuda, H., Ebata, A., Teramae, K. & Hishinuma, N. Alteration of Thermal Conductivity and Viscosity of Liquid by Dispersing Ultra-Fine Particles Dispersion of Al2O3, SiO2 and TiO2 Ultra-Fine Particles. *Netsu Bussei* **7**, 227-233 (1993).
13. S. U. S. Choi, D.A.S., H. P. Wang. 99-105 (New York, 1995).
14. Das, S.K. Nanofluids: Science and Technology (Wiley-Interscience, 2007).
15. Chandrasekar, M. & Suresh, S. A Review on the Mechanisms of Heat Transport in Nanofluids. *Heat Transfer Engineering* **30**, 1136-1150 (2009).
16. Özerinç, S., Kakaç, S. & Yazıcıoğlu, A. Enhanced thermal conductivity of nanofluids: a state-of-the-art review. *Microfluidics and Nanofluidics* **8**, 145-170 (2010).
17. Lee, J.H.L., S. H.; Choi, C. J.; Jang, S. P.; Choi, S. U. S. A Review of Thermal Conductivity Data, Mechanisms and Models for Nanofluids. *International Journal of Micro-Nano Scale Transport* **1**, 269 (2010).
18. Trisaksri, V. & Wongwises, S. Critical review of heat transfer characteristics of nanofluids. *Renewable and Sustainable Energy Reviews* **11**, 512-523 (2007).
19. Wang, X.-Q. & Mujumdar, A.S. Heat transfer characteristics of nanofluids: a review. *International Journal of Thermal Sciences* **46**, 1-19 (2007).
20. Cravalho, E.G., Tien, C.L. & Caren, R.P. Effect of Small Spacings on Radiative Transfer Between Two Dielectrics. *Journal of Heat Transfer* **89**, 351-358 (1967).
21. Polder, D. & Van Hove, M. Theory of Radiative Heat Transfer between Closely Spaced Bodies. *Physical Review B* **4**, 3303-3314 (1971).
22. Hargreaves, C.M. Anomalous radiative transfer between closely-spaced bodies. *Physics Letters A* **30**, 491-492 (1969).
23. Shen, S., Narayanaswamy, A. & Chen, G. Surface Phonon Polaritons Mediated Energy Transfer between Nanoscale Gaps. *Nano Lett* **9**, 2909-2913 (2009).
24. Yao, T. Thermal properties of AlAs/GaAs superlattices. *Applied Physics Letters* **51**, 1798-1800 (1987).
25. Chen, G., Tien, C.L., Wu, X. & Smith, J.S. Thermal Diffusivity Measurement of GaAs/AlGaAs Thin-Film Structures. *Journal of Heat Transfer* **116**, 325-331 (1994).
26. Lee, S.-M., Cahill, D.G. & Venkatasubramanian, R. Thermal conductivity of Si--Ge superlattices. *Applied Physics Letters* **70**, 2957-2959 (1997).
27. Borca-Tasciuc, T. et al. Thermal conductivity of symmetrically strained Si/Ge superlattices. *Superlattices and Microstructures* **28**, 199-206 (2000).
28. Li, D., Wu, Y., Fan, R., Yang, P. & Majumdar, A. Thermal conductivity of Si/SiGe superlattice nanowires. *Applied Physics Letters* **83**, 3186-3188 (2003).
29. Chen, G. in Semiconductors and Semimetals (ed. Terry, M.T.) 203-259 (Elsevier, 2001).
30. Li, D. et al. Thermal conductivity of individual silicon nanowires. *Applied Physics Letters* **83**, 2934 (2003).
31. Hochbaum, A.I. et al. Enhanced thermoelectric performance of rough silicon nanowires. *Nature* **451**, 163-U5 (2008).
32. Lim, J., Hippalgaonkar, K., Andrews, S.C., Majumdar, A. & Yang, P. Quantifying Surface Roughness Effects on Phonon Transport in Silicon Nanowires. *Nano Lett* **12**, 2475-2482 (2012).

33. Chiritescu, C. et al. Ultralow thermal conductivity in disordered, layered WSe(2) crystals. *Science* **315**, 351-353 (2007).
34. Poudel, B. et al. High-thermoelectric performance of nanostructured bismuth antimony telluride bulk alloys. *Science* **320**, 634-8 (2008).
35. Cahill, D.G. et al. Nanoscale thermal transport. *Journal of Applied Physics* **93**, 793-818 (2003).
36. Zhang, Z. Nano/Microscale Heat Transfer (McGraw-Hill, 2007).
37. Klemens, P.G. in Encyclopedia of Physics (ed. Flugge, S.) 198 (Springer, Berlin, 1956).
38. Callaway, J. Model for Lattice Thermal Conductivity at Low Temperatures. *Physical Review* **113**, 1046-1051 (1959).
39. Peierls, R. Zur kinetischen Theorie der Wärmeleitung in Kristallen. *Annalen der Physik* **395**, 1055-1101 (1929).
40. Klemens, P.G. & Pedraza, D.F. Thermal-Conductivity of Graphite in the Basal-Plane. *Carbon* **32**, 735-741 (1994).
41. Holland, M.G. Analysis of Lattice Thermal Conductivity. *Physical Review* **132**, 2461-2471 (1963).
42. Slack, G.A. & Galginaitis, S. Thermal Conductivity + Phonon Scattering by Magnetic Impurities in Cdte. *Physical Review* **133**, A253-A268 (1964).
43. Ward, A., Broido, D.A., Stewart, D.A. & Deinzer, G. Ab initio theory of the lattice thermal conductivity in diamond. *Physical Review B* **80** (2009).
44. Ward, A. & Broido, D.A. Intrinsic phonon relaxation times from first-principles studies of the thermal conductivities of Si and Ge. *Physical Review B* **81** (2010).
45. Esfarjani, K., Chen, G. & Stokes, H.T. Heat transport in silicon from first-principles calculations. *Physical Review B* **84**, 085204 (2011).
46. Yin, M.T. & Cohen, M.L. Theory of lattice-dynamical properties of solids: Application to Si and Ge. *Physical Review B* **26**, 3259-3272 (1982).
47. Giannozzi, P., De Gironcoli, S., Pavone, P. & Baroni, S. Abinitio Calculation of Phonon Dispersions in Semiconductors. *Physical Review B* **43**, 7231-7242 (1991).
48. DeCicco, P.D. & Johnson, F.A. The Quantum Theory of Lattice Dynamics. IV. *Proceedings of the Royal Society of London. A. Mathematical and Physical Sciences* **310**, 111-119 (1969).
49. Pick, R.M., Cohen, M.H. & Martin, R.M. Microscopic Theory of Force Constants in the Adiabatic Approximation. *Physical Review B* **1**, 910-920 (1970).
50. Gonze, X. & Vigneron, J.P. Density-Functional Approach to Nonlinear-Response Coefficients of Solids. *Physical Review B* **39**, 13120-13128 (1989).
51. Morse, P.M.C. & Feshbach, H. Methods of theoretical physics (McGraw-Hill, 1953).
52. Dalgarno, A. & Stewart, A.L. On the Perturbation Theory of Small Disturbances. *Proceedings of the Royal Society of London. Series A. Mathematical and Physical Sciences* **238**, 269-275 (1956).
53. King, H.F. & Komornicki, A. Analytic computation of energy derivatives. Relationships among partial derivatives of a variationally determined function. *The Journal of Chemical Physics* **84**, 5645-5650 (1986).
54. Debernardi, A., Baroni, S. & Molinari, E. Anharmonic Phonon Lifetimes in Semiconductors from Density-Functional Perturbation-Theory. *Physical Review Letters* **75**, 1819-1822 (1995).
55. Debernardi, A. Phonon linewidth in III-V semiconductors from density-functional perturbation theory. *Physical Review B* **57**, 12847-12858 (1998).
56. Broido, D.A., Malorny, M., Birner, G., Mingo, N. & Stewart, D.A. Intrinsic lattice thermal conductivity of semiconductors from first principles. *Applied Physics Letters* **91**, 231922 (2007).
57. Esfarjani, K. & Stokes, H.T. Method to extract anharmonic force constants from first principles calculations. *Physical Review B* **77** (2008).
58. Luo, T.F., Garg, J., Esfarjani, K., Shiomi, J. & Chen, G. Gallium Arsenide Thermal Conductivity and Optical Phonon Relaxation Times from First-Principles Calculations. *Europhys. Let.* **in press** (2012).
59. Shiomi, J., Esfarjani, K. & Chen, G. Thermal conductivity of half-Heusler compounds from first-principles calculations. *Physical Review B* **84**, 104302 (2011).
60. Shiga, T. et al. Microscopic mechanism of low thermal conductivity in lead telluride. *Physical Review B* **85**, 155203 (2012).
61. Tian, Z. et al. Phonon conduction in PbSe, PbTe, and PbTe$_{1-x}$Se$_{x}$ from first-principles calculations. *Physical Review B* **85**, 184303 (2012).
62. Esfarjani, K., Chen, G. & Stokes, H.T. Heat transport in silicon from first-principles calculations. *Physical Review B* **84** (2011).
63. Inyushkin, A.V., Taldenkov, A.N., Gibin, A.M., Gusev, A.V. & Pohl, H.J. On the isotope effect in thermal conductivity of silicon. *physica status solidi (c)* **1**, 2995-2998 (2004).
64. Inyushkin, A.V. et al. Thermal conductivity of isotopically enriched 71 GaAs crystal. *Semiconductor Science and Technology* **18**, 685 (2003).
65. Takeyuki Sekimoto, K.K., Hiroaki Muta, and Shinsuke Yamanaka High-Thermoelectric Figure of Merit Realized in p-Type Half-Heusler Compounds: ZrCoSnxSb1-x. *Jpn. J. Appl. Phys.* **46**, L673 (2007).
66. Xia, Y. et al. Thermoelectric properties of semimetallic (Zr, Hf)CoSb half-Heusler phases. *Journal of Applied Physics* **88**, 1952-1955 (2000).
67. Shiga, T. et al. Microscopic mechanism of low thermal conductivity in lead telluride. *Physical Review B* **85**, 155203 (2012).
68. Ravich, U.I., Stillbans, L.S. & Tybulewicz, A. Semiconducting Lead Chalcogenides (Books on Demand).
69. Tian, Z. et al. Phonon conduction in PbSe, PbTe, and PbTe$_{1-x}$Se$_{x}$ from first-principles calculations. *Physical Review B* **85**, 184303 (2012).
70. El-Sharkawy, A.A., Abou El-Azm, A.M., Kenawy, M.I., Hillal, A.S. & Abu-Basha, H.M. Thermophysical properties of polycrystalline PbS, PbSe, and PbTe in the temperature range 300–700 K. *International Journal of Thermophysics* **4**, 261-269 (1983).
71. Garg, J., Bonini, N., Kozinsky, B. & Marzari, N. Role of Disorder and Anharmonicity in the Thermal Conductivity of Silicon-Germanium Alloys: A First-Principles Study. *Physical Review Letters* **106** (2011).

72. Garg, J., Bonini, N. & Marzari, N. High Thermal Conductivity in Short-Period Superlattices. *Nano Lett* **11**, 5135-5141 (2011).
73. Lindsay, L. & Broido, D.A. Optimized Tersoff and Brenner empirical potential parameters for lattice dynamics and phonon thermal transport in carbon nanotubes and graphene (vol 81, 205441, 2010). *Physical Review B* **82** (2010).
74. Lindsay, L., Broido, D.A. & Mingo, N. Flexural phonons and thermal transport in graphene. *Physical Review B* **82**, 115427 (2010).
75. Lindsay, L., Broido, D.A. & Mingo, N. Flexural phonons and thermal transport in multilayer graphene and graphite. *Physical Review B* **83**, 235428 (2011).
76. Lindsay, L. & Broido, D.A. Enhanced thermal conductivity and isotope effect in single-layer hexagonal boron nitride. *Physical Review B* **84**, 155421 (2011).
77. MullerPlathe, F. A simple nonequilibrium molecular dynamics method for calculating the thermal conductivity. *Journal of Chemical Physics* **106**, 6082-6085 (1997).
78. Schelling, P.K., Phillpot, S.R. & Keblinski, P. Comparison of atomic-level simulation methods for computing thermal conductivity. *Physical Review B* **65** (2002).
79. Maruyama, S. in Advances in Numerical Heat Transfer (ed. W. J. Minkowycz, E.M.S.) (CRC Press, 2000).
80. Lukes, J.R., Li, D.Y., Liang, X.-G. & Tien, C.-L. Molecular Dynamics Study of Solid Thin-Film Thermal Conductivity. *Journal of Heat Transfer* **122**, 536-543 (2000).
81. Volz, S.G. & Chen, G. Molecular dynamics simulation of thermal conductivity of silicon nanowires. *Applied Physics Letters* **75**, 2056-2058 (1999).
82. Volz, S.G. & Chen, G. Molecular-dynamics simulation of thermal conductivity of silicon crystals. *Physical Review B* **61**, 2651-2656 (2000).
83. Kotake, S. & Wakuri, S. Molecular Dynamics Study of Heat Conduction in Solid Materials. *JSME international journal. Ser. B, Fluids and thermal engineering* **37**, 103-108 (1994).
84. Volz, S. et al. Transient Fourier-law deviation by molecular dynamics in solid argon. *Physical Review B* **54**, 340-347 (1996).
85. Minnich, A.J. et al. Thermal Conductivity Spectroscopy Technique to Measure Phonon Mean Free Paths. *Physical Review Letters* **107** (2011).
86. Li, J., Porter, L. & Yip, S. Atomistic modeling of finite-temperature properties of crystalline β-SiC: II. Thermal conductivity and effects of point defects. *Journal of Nuclear Materials* **255**, 139-152 (1998).
87. McGaughey, A. Thermal conductivity decomposition and analysis using molecular dynamics simulations Part II. Complex silica structures. *International Journal of Heat and Mass Transfer* **47**, 1799-1816 (2004).
88. Huang, B., McGaughey, A. & Kaviany, M. Thermal conductivity of metal-organic framework 5 (MOF-5): Part I. Molecular dynamics simulations. *International Journal of Heat and Mass Transfer* **50**, 393-404 (2007).
89. Qiu, B. & Ruan, X.L. Molecular dynamics simulations of lattice thermal conductivity of bismuth telluride using two-body interatomic potentials. *Physical Review B* **80** (2009).
90. McGaughey, A.J.H. & Kaviany, M. Observation and description of phonon interactions in molecular dynamics simulations. *Physical Review B* **71** (2005).
91. Henry, A.S. & Chen, G. Spectral phonon transport properties of silicon based on molecular dynamics Simulations and lattice dynamics. *Journal of Computational and Theoretical Nanoscience* **5**, 141-152 (2008).
92. Henry, A. & Chen, G. Anomalous heat conduction in polyethylene chains: Theory and molecular dynamics simulations. *Physical Review B* **79**, 144305 (2009).
93. Thomas, J.A., Turney, J.E., Iutzi, R.M., Amon, C.H. & McGaughey, A.J.H. Predicting phonon dispersion relations and lifetimes from the spectral energy density. *Physical Review B* **81** (2010).
94. Qiu, B., Bao, H., Zhang, G., Wu, Y. & Ruan, X. Molecular dynamics simulations of lattice thermal conductivity and spectral phonon mean free path of PbTe: Bulk and nanostructures. *Computational Materials Science* **53**, 278-285 (2012).
95. de Koker, N. Thermal Conductivity of MgO Periclase from Equilibrium First Principles Molecular Dynamics. *Physical Review Letters* **103**, 125902 (2009).
96. Luo, T. & Lloyd, J.R. Ab Initio Molecular Dynamics Study of Nanoscale Thermal Energy Transport. *Journal of Heat Transfer* **130**, 122403 (2008).
97. Huang, B.L., McGaughey, A.J.H. & Kaviany, M. Thermal conductivity of metal-organic framework 5 (MOF-5): Part I. Molecular dynamics simulations. *International Journal of Heat and Mass Transfer* **50**, 393-404 (2007).
98. Qiu, B., Sun, L. & Ruan, X.L. Lattice thermal conductivity reduction in Bi(2)Te(3) quantum wires with smooth and rough surfaces: A molecular dynamics study. *Physical Review B* **83** (2011).
99. Huang, B.-L. & Kaviany, M. Ab initio and molecular dynamics predictions for electron and phonon transport in bismuth telluride. *Physical Review B* **77**, 125209 (2008).
100. Broido, D.A., Ward, A. & Mingo, N. Lattice thermal conductivity of silicon from empirical interatomic potentials. *Physical Review B* **72** (2005).
101. Garg, J. Thermal Conductivity from First-Principle in Bulk, Disordered, and Nanostructured Materials. *Ph.D. Thesis, Department of Mechanical Engineering, MIT* (2011).
102. Schelling, P.K., Phillpot, S.R. & Keblinski, P. Phonon wave-packet dynamics at semiconductor interfaces by molecular-dynamics simulation. *Applied Physics Letters* **80**, 2484 (2002).
103. Hu, M., Keblinski, P. & Schelling, P. Kapitza conductance of silicon–amorphous polyethylene interfaces by molecular dynamics simulations. *Physical Review B* **79** (2009).
104. Tian, Z.T., Kim, S., Sun, Y. & White, B. A Molecular Dynamics Study of Thermal Conductivity in Nanocomposites Via the Phonon Wave Packet Method. *Ipack 2009: Proceedings of the Asme Interpack Conference 2009, Vol 1*, 607-615 (2010).
105. Tian, Z.T., White, B.E. & Sun, Y. Phonon wave-packet interference and phonon tunneling based energy transport across nanostructured thin films. *Applied Physics Letters* **96** (2010).
106. Kumar, S. & Murthy, J.Y. Interfacial thermal transport between nanotubes. *Journal of Applied Physics* **106** (2009).

107. Chen, L. & Kumar, S. Thermal transport in double-wall carbon nanotubes using heat pulse. *Journal of Applied Physics* **110** (2011).
108. Shenogin, S. Role of thermal boundary resistance on the heat flow in carbon-nanotube composites. *Journal of Applied Physics* **95**, 8136 (2004).
109. Patel, H.A., Garde, S. & Keblinski, P. Thermal resistance of nanoscopic liquid-liquid interfaces: Dependence on chemistry and molecular architecture. *Nano Lett* **5**, 2225-2231 (2005).
110. Luo, T. & Lloyd, J.R. Non-equilibrium molecular dynamics study of thermal energy transport in Au–SAM–Au junctions. *International Journal of Heat and Mass Transfer* **53**, 1-11 (2010).
111. Luo, T. & Lloyd, J.R. Molecular dynamics study of thermal transport in GaAs-self-assembly monolayer-GaAs junctions with ab initio characterization of thiol-GaAs bonds. *Journal of Applied Physics* **109**, 034301 (2011).
112. Luo, T. & Lloyd, J.R. Equilibrium Molecular Dynamics Study of Lattice Thermal Conductivity/Conductance of Au-SAM-Au Junctions. *Journal of Heat Transfer* **132**, 032401 (2010).
113. Carlborg, C., Shiomi, J. & Maruyama, S. Thermal boundary resistance between single-walled carbon nanotubes and surrounding matrices. *Physical Review B* **78** (2008).
114. Huxtable, S.T. et al. Interfacial heat flow in carbon nanotube suspensions. *Nat Mater* **2**, 731-4 (2003).
115. Luo, T.F. & Lloyd, J.R. Enhancement of Thermal Energy Transport across Graphene/Graphite and Polymer Interfaces - A Molecular Dynamics Study. *Advanced Functional Materials* **22**, 2495 (2012).
116. Hu, L., Desai, T. & Keblinski, P. Thermal transport in graphene-based nanocomposite. *Journal of Applied Physics* **110** (2011).
117. Shenogina, N., Keblinski, P. & Garde, S. Strong frequency dependence of dynamical coupling between protein and water. *Journal of Chemical Physics* **129** (2008).
118. Chenoweth, K., Cheung, S., van Duin, A.C., Goddard, W.A., 3rd & Kober, E.M. Simulations on the thermal decomposition of a poly(dimethylsiloxane) polymer using the ReaxFF reactive force field. *J Am Chem Soc* **127**, 7192-202 (2005).
119. Stuart, S.J., Tutein, A.B. & Harrison, J.A. A reactive potential for hydrocarbons with intermolecular interactions. *The Journal of Chemical Physics* **112**, 6472-6486 (2000).
120. Mingo, N. & Yang, L. Phonon transport in nanowires coated with an amorphous material: An atomistic Green's function approach. *Physical Review B* **68** (2003).
121. Economou, E.N. Green's Functions in Quantum Physics (Springer-Verlag, Berlin, 1983).
122. Mingo, N. Anharmonic phonon flow through molecular-sized junctions. *Physical Review B* **74**, 125402 (2006).
123. Wang, J.S., Wang, J. & Zeng, N. Nonequilibrium Green's function approach to mesoscopic thermal transport. *Physical Review B* **74** (2006).
124. Hopkins, P.E. & Serrano, J.R. Phonon localization and thermal rectification in asymmetric harmonic chains using a nonequilibrium Green's function formalism. *Physical Review B* **80** (2009).
125. Zhang, W., Fisher, T.S. & Mingo, N. Simulation of interfacial phonon transport in Si-Ge heterostructures using an atomistic Green's function method. *Journal of Heat Transfer-Transactions of the Asme* **129**, 483-491 (2007).
126. Huang, Z., Fisher, T. & Murthy, J. An atomistic study of thermal conductance across a metal-graphene nanoribbon interface. *Journal of Applied Physics* **109**, 074305 (2011).
127. Huang, Z., Fisher, T.S. & Murthy, J.Y. Simulation of phonon transmission through graphene and graphene nanoribbons with a Green's function method. *Journal of Applied Physics* **108**, 094319 (2010).
128. Ziman, J.M. Electrons and Phonons: The Theory of Transport Phenomena in Solids (Oxford University Press, USA, 2001).
129. Murthy, J.Y.N., S. V. J.; Pascual-Gutierrez, J. A.; Wang, T.; Ni, C.; Mathur, S. R. Review of Multi-Scale Simulation in Sub-Micron Heat Transfer *International Journal for Multiscale Computational Engineering* **3**, 135 (2005).
130. Minnich, A.J., Chen, G., Mansoor, S. & Yilbas, B.S. Quasiballistic heat transfer studied using the frequency-dependent Boltzmann transport equation. *Physical Review B* **84**, 235207 (2011).
131. Tian, W. & Yang, R. Thermal conductivity modeling of compacted nanowire composites. *Journal of Applied Physics* **101**, 054320 (2007).
132. Jeng, M.-S., Yang, R., Song, D. & Chen, G. Modeling the Thermal Conductivity and Phonon Transport in Nanoparticle Composites Using Monte Carlo Simulation. *Journal of Heat Transfer* **130**, 042410 (2008).
133. Klitsner, T., VanCleve, J.E., Fischer, H.E. & Pohl, R.O. Phonon radiative heat transfer and surface scattering. *Physical Review B* **38**, 7576-7594 (1988).
134. Peterson, R.B. Direct Simulation of Phonon-Mediated Heat Transfer in a Debye Crystal. *Journal of Heat Transfer* **116**, 815-822 (1994).
135. Hao, Q., Chen, G. & Jeng, M.-S. Frequency-dependent Monte Carlo simulations of phonon transport in two-dimensional porous silicon with aligned pores. *Journal of Applied Physics* **106**, 114321 (2009).
136. Mazumder, S. & Majumdar, A. Monte Carlo Study of Phonon Transport in Solid Thin Films Including Dispersion and Polarization. *Journal of Heat Transfer* **123**, 749-759 (2001).
137. Péraud, J.-P.M. & Hadjiconstantinou, N.G. Efficient simulation of multidimensional phonon transport using energy-based variance-reduced Monte Carlo formulations. *Physical Review B* **84**, 205331 (2011).
138. Peraud, J.-P.M. & Hadjiconstantinou, N.G. An alternative approach to efficient simulation of micro/nanoscale phonon transport. *Applied Physics Letters* **101**, 153114 (2012).
139. McGaughey, A.J.H. & Jain, A. Nanostructure thermal conductivity prediction by Monte Carlo sampling of phonon free paths. *Applied Physics Letters* **100**, 061911 (2012).
140. Sofo, J.O. & Mahan, G.D. Diffusion and transport coefficients in synthetic opals. *Physical Review B* **62**, 2780-2785 (2000).
141. Maxwell, J.C. & Thompson, J.J. A Treatise on Electricity and Magnetism (Clarendon, 1904).

142. Hasselman, D.P.H. & Johnson, L.F. Effective Thermal Conductivity of Composites with Interfacial Thermal Barrier Resistance. *Journal of Composite Materials* **21**, 508-515 (1987).
143. Benveniste, Y. Effective thermal conductivity of composites with a thermal contact resistance between the constituents: Nondilute case. *Journal of Applied Physics* **61**, 2840-2843 (1987).
144. Yang, R., Chen, G. & Dresselhaus, M.S. Thermal Conductivity Modeling of Core−Shell and Tubular Nanowires. *Nano Lett* **5**, 1111-1115 (2005).
145. Minnich, A. & Chen, G. Modified effective medium formulation for the thermal conductivity of nanocomposites. *Applied Physics Letters* **91**, 073105 (2007).
146. Cahill, D.G. Thermal conductivity measurement from 30 to 750 K: the 3 omega method. *Review of Scientific Instruments* **61**, 802-808 (1990).
147. Birge, N.O. Specific-heat spectroscopy of glycerol and propylene glycol near the glass transition. *Physical Review B* **34**, 1631-1642 (1986).
148. Cahill, D.G. Thermal-Conductivity Measurement from 30-K to 750-K - the 3-Omega Method. *Review of Scientific Instruments* **61**, 802-808 (1990).
149. Cahill, D.G., Katiyar, M. & Abelson, J.R. Thermal-Conductivity of Alpha-Sih Thin-Films. *Physical Review B* **50**, 6077-6081 (1994).
150. Borca-Tasciuc, T., Kumar, A.R. & Chen, G. Data reduction in 3 omega method for thin-film thermal conductivity determination. *Review of Scientific Instruments* **72**, 2139-2147 (2001).
151. Lu, L., Yi, W. & Zhang, D.L. 3 omega method for specific heat and thermal conductivity measurements. *Review of Scientific Instruments* **72**, 2996-3003 (2001).
152. Choi, T.-Y., Poulikakos, D., Tharian, J. & Sennhauser, U. Measurement of the Thermal Conductivity of Individual Carbon Nanotubes by the Four-Point Three-ω Method. *Nano Lett* **6**, 1589-1593 (2006).
153. Birge, N.O. & Nagel, S.R. Wide-frequency specific heat spectrometer. *Review of Scientific Instruments* **58**, 1464-1470 (1987).
154. Wang, R.Y., Segalman, R.A. & Majumdar, A. Room temperature thermal conductance of alkanedithiol self-assembled monolayers. *Applied Physics Letters* **89**, 173113 (2006).
155. Hu, X.J., Padilla, A.A., Xu, J., Fisher, T.S. & Goodson, K.E. 3-Omega Measurements of Vertically Oriented Carbon Nanotubes on Silicon. *Journal of Heat Transfer* **128**, 1109-1113 (2006).
156. Yang, B., Liu, W.L., Liu, J.L., Wang, K.L. & Chen, G. Measurements of anisotropic thermoelectric properties in superlattices. *Applied Physics Letters* **81**, 3588-3590 (2002).
157. Dames, C. & Chen, G. 1 omega, 2 omega, and 3 omega methods for measurements of thermal properties. *Review of Scientific Instruments* **76**, 124902 (2005).
158. Borca-Tasciuc, T. & Chen, G. in Thermal Conductivity: Theory, Properties, and Applications (ed. Tritt, T.M.) (Kluwar Press, New York, 2004).
159. Tai, Y.-C., Hsu, T.-Y. & Hsieh, W.H. (ed. Patent, U.S.) (United States, 2001).
160. Volklein, F. & Balles, H. A Microstructure For Measurement Of Thermal Conductivity Of Polysilicon Thin Films. *Microelectromechanical Systems, Journal of* **1**, 193-196 (1992).
161. Jain, A. & Goodson, K.E. Measurement of the Thermal Conductivity and Heat Capacity of Freestanding Shape Memory Thin Films Using the 3 omega Method. *Journal of Heat Transfer* **130**, 102402 (2008).
162. Goodson, K.E. & Ju, Y.S. HEAT CONDUCTION IN NOVEL ELECTRONIC FILMS. *Annual Review of Materials Science* **29**, 261-293 (1999).
163. Kim, P., Shi, L., Majumdar, A. & McEuen, P.L. Thermal Transport Measurements of Individual Multiwalled Nanotubes. *Physical Review Letters* **87**, 215502 (2001).
164. Shi, L. et al. Measuring Thermal and Thermoelectric Properties of One-Dimensional Nanostructures Using a Microfabricated Device. *Journal of Heat Transfer* **125**, 881-888 (2003).
165. Pettes, M.T. & Shi, L. Thermal and Structural Characterizations of Individual Single-, Double-, and Multi-Walled Carbon Nanotubes. *Advanced Functional Materials* **19**, 3918-3925 (2009).
166. Zhou, F. et al. Thermal conductivity of indium arsenide nanowires with wurtzite and zinc blende phases. *Physical Review B* **83**, 205416 (2011).
167. Moore, A.L., Pettes, M.T., Zhou, F. & Shi, L. Thermal conductivity suppression in bismuth nanowires. *Journal of Applied Physics* **106**, 034310 (2009).
168. Seol, J.H. et al. Two-Dimensional Phonon Transport in Supported Graphene. *Science* **328**, 213-216 (2010).
169. Yang, J. et al. Enhanced and switchable nanoscale thermal conduction due to van der Waals interfaces. *Nat Nano* **7**, 91-95 (2012).
170. Yang, J. et al. Contact thermal resistance between individual multiwall carbon nanotubes. *Applied Physics Letters* **96**, 023109 (2010).
171. Fujii, M. et al. Measuring the Thermal Conductivity of a Single Carbon Nanotube. *Physical Review Letters* **95**, 065502 (2005).
172. Harris, C.T. et al. Fabrication of a nanostructure thermal property measurement platform. *Nanotechnology* **22**, 275308 (2011).
173. Dames, C. et al. A hot-wire probe for thermal measurements of nanowires and nanotubes inside a transmission electron microscope. *Review of Scientific Instruments* **78**, 104903 (2007).
174. Williams, C.C. & Wickramasinghe, H.K. Scanning thermal profiler. *Applied Physics Letters* **49**, 1587-1589 (1986).
175. Binnig, G., Quate, C.F. & Gerber, C. Atomic Force Microscope. *Physical Review Letters* **56**, 930-933 (1986).
176. Pylkki, R. et al. Scanning Near-Field Optical Microscopy and Scanning Thermal Microscopy. *Japanese Journal of Applied Physics* **33**, 3785.
177. Suzuki, Y. Novel Microcantilever for Scanning Thermal Imaging Microscopy. *Japanese Journal of Applied Physics* **35**, L352.
178. Luo, K., Shi, Z., Lai, J. & Majumdar, A. Nanofabrication of sensors on cantilever probe tips for scanning multiprobe microscopy. *Applied Physics Letters* **68**, 325-327 (1996).

179. Hinz, M., Marti, O., Gotsmann, B., Lantz, M.A. & Durig, U. High resolution vacuum scanning thermal microscopy of HfO[sub 2] and SiO[sub 2]. *Applied Physics Letters* **92**, 043122 (2008).
180. Mills, G., Zhou, H., Midha, A., Donaldson, L. & Weaver, J.M.R. Scanning thermal microscopy using batch fabricated thermocouple probes. *Applied Physics Letters* **72**, 2900-2902 (1998).
181. Li, M.H., Wu, J.J. & Gianchandani, Y.B. Surface micromachined polyimide scanning thermocouple probes. *Microelectromechanical Systems, Journal of* **10**, 3-9 (2001).
182. Hammiche, A., Pollock, H.M., Song, M. & Hourston, D.J. Sub-surface imaging by scanning thermal microscopy. *Measurement Science and Technology* **7**, 142 (1996).
183. Ruiz, F., Sun, W.D., Pollak, F.H. & Venkatraman, C. Determination of the thermal conductivity of diamond-like nanocomposite films using a scanning thermal microscope. *Applied Physics Letters* **73**, 1802-1804 (1998).
184. Majumdar, A. Scanning thermal microscopy. *Annual Review of Materials Science* **29**, 505-585 (1999).
185. Kim, K., Jeong, W., Lee, W. & Reddy, P. Ultra-High Vacuum Scanning Thermal Microscopy for Nanometer Resolution Quantitative Thermometry. *Acs Nano* **6**, 4248-4257 (2012).
186. Kim, K. et al. Quantitative scanning thermal microscopy using double scan technique. *Applied Physics Letters* **93**, 203115 (2008).
187. Menges, F., Riel, H., Stemmer, A. & Gotsmann, B. Quantitative Thermometry of Nanoscale Hot Spots. *Nano Lett* **12**, 596-601 (2012).
188. Cretin, B., Gomès, S., Trannoy, N. & Vairac, P. (ed. Volz, S.) 181-238 (Springer Berlin / Heidelberg, 2007).
189. Igeta, M., Inoue, T., Varesi, J. & Majumdar, A. Thermal Expansion and Temperature Measurement in a Microscopic Scale by Using the Atomic Force Microscope. *JSME International Journal Series B* **42**, 723-730 (1999).
190. Nakabeppu, O., Chandrachood, M., Wu, Y., Lai, J. & Majumdar, A. Scanning thermal imaging microscopy using composite cantilever probes. *Applied Physics Letters* **66**, 694-696 (1995).
191. Shi, L. & Majumdar, A. Thermal Transport Mechanisms at Nanoscale Point Contacts. *Journal of Heat Transfer* **124**, 329-337 (2002).
192. Gimzewski, J.K., Gerber, C., Meyer, E. & Schlittler, R.R. Observation of a chemical reaction using a micromechanical sensor. *Chemical Physics Letters* **217**, 589-594 (1994).
193. Barnes, J.R., Stephenson, R.J., Welland, M.E., Gerber, C. & Gimzewski, J.K. Photothermal spectroscopy with femtojoule sensitivity using a micromechanical device. *Nature* **372**, 79-81 (1994).
194. Sulchek, T. et al. Dual integrated actuators for extended range high speed atomic force microscopy. *Applied Physics Letters* **75**, 1637-1639 (1999).
195. Lai, J., Perazzo, T., Shi, Z. & Majumdar, A. Optimization and performance of high-resolution micro-optomechanical thermal sensors. *Sensors and Actuators A: Physical* **58**, 113-119 (1997).
196. Sadat, S. et al. Room temperature picowatt-resolution calorimetry. *Applied Physics Letters* **99**, 043106 (2011).
197. Eesley, G.L. Observation of Nonequilibrium Electron Heating in Copper. *Physical Review Letters* **51**, 2140-2143 (1983).
198. Opsal, J., Taylor, M.W., Smith, W.L. & Rosencwaig, A. Temporal behavior of modulated optical reflectance in silicon. *Journal of Applied Physics* **61**, 240-248 (1987).
199. Paddock, C.A. & Eesley, G.L. Transient Thermoreflectance from Thin Metal-Films. *Journal of Applied Physics* **60**, 285-290 (1986).
200. Chen, G. & Borca-Tasciuc, T. Applicability of photothermal radiometry for temperature measurement of semiconductors. *International Journal of Heat and Mass Transfer* **41**, 2279-2285 (1998).
201. Loarer, T., Greffet, J.-J. & Huetz-Aubert, M. Noncontact surface temperature measurement by means of a modulated photothermal effect. *Appl. Opt.* **29**, 979-987 (1990).
202. Markham, J.R. et al. Bench top Fourier transform infrared based instrument for simultaneously measuring surface spectral emittance and temperature. *Review of Scientific Instruments* **64**, 2515-2522 (1993).
203. Loarer, T. & Greffet, J.-J. Application of the pulsed photothermal effect to fast surface temperature measurements. *Appl. Opt.* **31**, 5350-5358 (1992).
204. Gao, B. et al. Studies of Intrinsic Hot Phonon Dynamics in Suspended Graphene by Transient Absorption Microscopy. *Nano Lett* **11**, 3184-3189 (2011).
205. Balderas-Lopez, J.A., Mandelis, A. & Garcia, J.A. Normalized photoacoustic techniques for thermal diffusivity measurements of buried layers in multilayered systems. *Journal of Applied Physics* **92**, 3047-3055 (2002).
206. Nelson, K.A., Casalegno, R., Miller, R.J.D. & Fayer, M.D. Laser-induced excited state and ultrasonic wave gratings: Amplitude and phase grating contributions to diffraction. *The Journal of Chemical Physics* **77**, 1144-1152 (1982).
207. Johnson, J.A. Optical Characterization of Complex Mechanical and Thermal Transport Properties (Ph.D. Thesis, Department of Chemistry, MIT, 2011).
208. Eichler, H., Salje, G. & Stahl, H. Thermal diffusion measurements using spatially periodic temperature distributions induced by laser light. *Journal of Applied Physics* **44**, 5383-5388 (1973).
209. Mandelis, A. Photothermal applications to the thermal analysis of solids. *Journal of Thermal Analysis and Calorimetry* **37**, 1065-1101 (1991).
210. Rosencwaig, A., Opsal, J., Smith, W.L. & Willenborg, D.L. Detection of thermal waves through optical reflectance. *Applied Physics Letters* **46**, 1013-1015 (1985).
211. Cahill, D.G. Analysis of heat flow in layered structures for time-domain thermoreflectance. *Review of Scientific Instruments* **75**, 5119-5122 (2004).
212. Schmidt, A.J. Optical Characterization of Thermal Transport from the Nanoscale to the Macroscale (Ph.D. Thesis, Department of Mechanical Engineering, MIT, 2008).
213. Norris, P.M. et al. Femtosecond pump--probe nondestructive examination of materials (invited). *Review of Scientific Instruments* **74**, 400-406 (2003).

214. Koh, Y.K. & Cahill, D.G. Frequency dependence of the thermal conductivity of semiconductor alloys. *Physical Review B* **76** (2007).
215. Rogers, J.A., Maznev, A.A., Banet, M.J. & Nelson, K.A. OPTICAL GENERATION AND CHARACTERIZATION OF ACOUSTIC WAVES IN THIN FILMS: Fundamentals and Applications. *Annual Review of Materials Science* **30**, 117-157 (2000).
216. Schmidt, A.J. et al. Probing the Gold Nanorod−Ligand−Solvent Interface by Plasmonic Absorption and Thermal Decay. *The Journal of Physical Chemistry C* **112**, 13320-13323 (2008).
217. Paddock, C.A. & Eesley, G.L. Transient thermoreflectance from thin metal films. *Journal of Applied Physics* **60**, 285-290 (1986).
218. Clemens, B.M., Eesley, G.L. & Paddock, C.A. Time-resolved thermal transport in compositionally modulated metal films. *Physical Review B* **37**, 1085-1096 (1988).
219. Capinski, W.S. & Maris, H.J. Improved apparatus for picosecond pump-and-probe optical measurements. *Review of Scientific Instruments* **67**, 2720-2726 (1996).
220. Costescu, R.M., Wall, M.A. & Cahill, D.G. Thermal conductance of epitaxial interfaces. *Physical Review B* **67** (2003).
221. Lyeo, H.K. & Cahill, D.G. Thermal conductance of interfaces between highly dissimilar materials. *Physical Review B* **73** (2006).
222. Gundrum, B.C., Cahill, D.G. & Averback, R.S. Thermal conductance of metal-metal interfaces. *Physical Review B* **72** (2005).
223. Cahill, D.G. & Watanabe, F. Thermal conductivity of isotopically pure and Ge-doped Si epitaxial layers from 300 to 550 K. *Physical Review B* **70** (2004).
224. Schmidt, A.J., Collins, K.C., Minnich, A.J. & Chen, G. Thermal conductance and phonon transmissivity of metal-graphite interfaces. *Journal of Applied Physics* **107** (2010).
225. Malen, J.A. et al. Optical Measurement of Thermal Conductivity Using Fiber Aligned Frequency Domain Thermoreflectance. *Journal of Heat Transfer* **133**, 081601 (2011).
226. Boccara, A.C., Fournier, D. & Badoz, J. Thermo-optical spectroscopy: Detection by the "mirage effect". *Applied Physics Letters* **36**, 130-132 (1980).
227. Murphy, J.C. & Aamodt, L.C. Photothermal spectroscopy using optical beam probing: Mirage effect. *Journal of Applied Physics* **51**, 4580-4588 (1980).
228. Cai, W. et al. Thermal Transport in Suspended and Supported Monolayer Graphene Grown by Chemical Vapor Deposition. *Nano Lett* **10**, 1645-1651 (2010).
229. Westover, T. et al. Photoluminescence, Thermal Transport, and Breakdown in Joule-Heated GaN Nanowires. *Nano Lett* **9**, 257-263 (2008).
230. Mahan, G.D. & Claro, F. Nonlocal theory of thermal conductivity. *Physical Review B* **38**, 1963-1969 (1988).
231. Chen, G. Nonlocal and Nonequilibrium Heat Conduction in the Vicinity of Nanoparticles. *Journal of Heat Transfer* **118**, 539-545 (1996).
232. Volz, S.G. Thermal insulating behavior in crystals at high frequencies. *Physical Review Letters* **87** (2001).
233. Carslaw, H.S. & Jaeger, J.C. Conduction of Heat in Solids (Oxford University Press, Oxford, 1959).
234. Siemens, M.E. et al. Quasi-ballistic thermal transport from nanoscale interfaces observed using ultrafast coherent soft X-ray beams. *Nat Mater* **9**, 26-30 (2010).
235. Hippalgaonkar, K. et al. Fabrication of Microdevices with Integrated Nanowires for Investigating Low-Dimensional Phonon Transport. *Nano Lett* **10**, 4341-4348 (2010).
236. Mingo, N., Yang, L., Li, D. & Majumdar, A. Predicting the Thermal Conductivity of Si and Ge Nanowires. *Nano Lett* **3**, 1713-1716 (2003).
237. Tian, Z., Esfarjani, K., Shiomi, J., Henry, A.S. & Chen, G. On the importance of optical phonons to thermal conductivity in nanostructures. *Applied Physics Letters* **99**, 053122-3 (2011).
238. Donadio, D. & Galli, G. Atomistic Simulations of Heat Transport in Silicon Nanowires. *Physical Review Letters* **102**, 195901 (2009).
239. Boukai, A.I. et al. Silicon nanowires as efficient thermoelectric materials. *Nature* **451**, 168-171 (2008).
240. Murphy, P.G. & Moore, J.E. Coherent phonon scattering effects on thermal transport in thin semiconductor nanowires. *Physical Review B* **76**, 155313 (2007).
241. Chen, R. et al. Thermal Conductance of Thin Silicon Nanowires. *Physical Review Letters* **101**, 105501 (2008).
242. Chu, D., Touzelbaev, M., Goodson, K.E., Babin, S. & Pease, R.F. 2874-2877 (AVS, 2001).
243. Touzelbaev, M.N. & Goodson, K.E. Impact of Experimental Timescale and Geometry on Thin-Film Thermal Property Measurements. *International Journal of Thermophysics* **22**, 243-263 (2001).
244. Asheghi, M., Leung, Y.K., Wong, S.S. & Goodson, K.E. Phonon-boundary scattering in thin silicon layers. *Applied Physics Letters* **71**, 1798-1800 (1997).
245. Panzer, M.A. et al. Thermal Properties of Ultrathin Hafnium Oxide Gate Dielectric Films. *Electron Device Letters, IEEE* **30**, 1269-1271 (2009).
246. Song, D.W. et al. Thermal conductivity of skutterudite thin films and superlattices. *Applied Physics Letters* **77**, 3854-3856 (2000).
247. Chen, G. Size and Interface Effects on Thermal Conductivity of Superlattices and Periodic Thin-Film Structures. *Journal of Heat Transfer* **119**, 220-229 (1997).
248. Goodson, K.E. Thermal Conduction in Nonhomogeneous CVD Diamond Layers in Electronic Microstructures. *Journal of Heat Transfer* **118**, 279-286 (1996).
249. Majumdar, A. Microscale Heat Conduction in Dielectric Thin Films. *Journal of Heat Transfer* **115**, 7-16 (1993).
250. Casimir, H.B.G. Note on the conduction of heat in crystals. *Physica* **5**, 495-500 (1938).
251. Fuchs, K. in Proc. Cambridge Philos. Soc 100 (Cambridge Univ Press, 1938).
252. Sondheimer, E.H. The mean free path of electrons in metals. *Advances in Physics* **50**, 499-537 (2001).
253. Lee, S.M., Cahill, D.G. & Venkatasubramanian, R. Thermal conductivity of Si-Ge superlattices. *Applied Physics Letters* **70**, 2957-2959 (1997).
254. Dames, C. & Chen, G. Theoretical phonon thermal conductivity of Si/Ge superlattice nanowires. *Journal of Applied Physics* **95**, 682-693 (2004).

255. Yang, B. & Chen, G. Partially coherent phonon heat conduction in superlattices. *Physical Review B* **67**, 195311 (2003).
256. Harman, T.C., Taylor, P.J., Walsh, M.P. & LaForge, B.E. Quantum dot superlattice thermoelectric materials and devices. *Science* **297**, 2229-2232 (2002).
257. Venkatasubramanian, R., Siivola, E., Colpitts, T. & O'Quinn, B. Thin-film thermoelectric devices with high room-temperature figures of merit. *Nature* **413**, 597-602 (2001).
258. Tamura, S., Tanaka, Y. & Maris, H.J. Phonon group velocity and thermal conduction in superlattices. *Physical Review B* **60**, 2627-2630 (1999).
259. Tellier, C.R. & Tosser, A.J. Size effects in thin films (Elsevier, 1982).
260. Chen, G. Thermal conductivity and ballistic-phonon transport in the cross-plane direction of superlattices. *Physical Review B* **57**, 14958-14973 (1998).
261. Luckyanova, M.N. et al. Coherent Phonon Heat Conduction in Superlattices. *Science* **338**, 936-939 (2012).
262. Chiritescu, C. et al. Ultralow Thermal Conductivity in Disordered, Layered WSe2 Crystals. *Science* **315**, 351-353 (2007).
263. Ford, J. The Fermi-Pasta-Ulam problem: paradox turns discovery. *Physics Reports* **213**, 271-310 (1992).
264. Li, B., Wang, J., Wang, L. & Zhang, G. Anomalous heat conduction and anomalous diffusion in nonlinear lattices, single walled nanotubes, and billiard gas channels. *Chaos: An Interdisciplinary Journal of Nonlinear Science* **15**, 015121 (2005).
265. Yang, N., Zhang, G. & Li, B. Violation of Fourier's law and anomalous heat diffusion in silicon nanowires. *Nano Today* **5**, 85-90 (2010).
266. Shiomi, J. & Maruyama, S. Diffusive-Ballistic Heat Conduction of Carbon Nanotubes and Nanographene Ribbons. *International Journal of Thermophysics* **31**, 1945-1951 (2010).
267. Klemens, P.G. Theory of the A-Plane Thermal Conductivity of Graphite. *Journal of Wide Bandgap Materials* **7**, 332-339 (2000).
268. Nika, D.L., Pokatilov, E.P., Askerov, A.S. & Balandin, A.A. Phonon thermal conduction in graphene: Role of Umklapp and edge roughness scattering. *Physical Review B* **79**, 155413 (2009).
269. Ghosh, S. et al. Extremely high thermal conductivity of graphene: Prospects for thermal management applications in nanoelectronic circuits. *Applied Physics Letters* **92** (2008).
270. Yan, Z., Liu, G., Khan, J.M. & Balandin, A.A. Graphene quilts for thermal management of high-power GaN transistors. *Nat Commun* **3**, 827 (2012).
271. Chen, Z., Jang, W., Bao, W., Lau, C.N. & Dames, C. Thermal contact resistance between graphene and silicon dioxide. *Applied Physics Letters* **95**, 161910 (2009).
272. Koh, Y.K., Bae, M.H., Cahill, D.G. & Pop, E. Heat Conduction across Monolayer and Few-Layer Graphenes. *Nano Letters* **10**, 4363-4368 (2010).
273. Savvides, N. & Goldsmid, H.J. The effect of boundary scattering on the high-temperature thermal conductivity of silicon. *Journal of Physics C: Solid State Physics* **6**, 1701 (1973).
274. Girard, S.N. et al. In Situ Nanostructure Generation and Evolution within a Bulk Thermoelectric Material to Reduce Lattice Thermal Conductivity. *Nano Lett* **10**, 2825-2831 (2010).
275. Allen, P., Feldman, J., Fabian, J. & Wooten, F. Diffusons, locons and propagons: Character of atomie yibrations in amorphous Si. *Philosophical Magazine Part B* **79**, 1715-1731 (1999).
276. Fujishiro, H., Ikebe, M., Kashima, T. & Yamanaka, A. Thermal conductivity and diffusivity of high-strength polymer fibers. *Japanese Journal of Applied Physics Part 1-Regular Papers Short Notes & Review Papers* **36**, 5633-5637 (1997).
277. Henry, A., Chen, G., Plimpton, S.J. & Thompson, A. 1D-to-3D transition of phonon heat conduction in polyethylene using molecular dynamics simulations. *Physical Review B* **82**, 144308 (2010).
278. Luo, T.F., Esfarjani, K., Shiomi, J., Henry, A. & Chen, G. Molecular dynamics simulation of thermal energy transport in polydimethylsiloxane (PDMS). *Journal of Applied Physics* **109** (2011).
279. Zhang, T. & Luo, T. Morphology-Influenced Thermal Conductivity of Polyethylene Single Chains and Crystalline Fibers. *Journal of Applied Physics* **112**, 094304 (2012).
280. Liu, J. & Yang, R. Length-dependent thermal conductivity of single extended polymer chains. *Physical Review B* **86**, 104307 (2012).
281. Eapen, J., Rusconi, R., Piazza, R. & Yip, S. The Classical Nature of Thermal Conduction in Nanofluids. *Journal of Heat Transfer* **132**, 102402 (2010).
282. Wang, J.J., Zheng, R.T., Gao, J.W. & Chen, G. Heat conduction mechanisms in nanofluids and suspensions. *Nano Today* **7**, 124-136 (2012).
283. Lu, P.J. et al. Gelation of particles with short-range attraction. *Nature* **453**, 499-503 (2008).
284. Lu, P.J., Conrad, J.C., Wyss, H.M., Schofield, A.B. & Weitz, D.A. Fluids of Clusters in Attractive Colloids. *Physical Review Letters* **96**, 028306 (2006).
285. Emanuela, Z., Peter, J.L., Fabio, C., David, A.W. & Francesco, S. Gelation as arrested phase separation in short-ranged attractive colloid–polymer mixtures. *Journal of Physics: Condensed Matter* **20**, 494242 (2008).
286. Gao, J.W., Zheng, R.T., Ohtani, H., Zhu, D.S. & Chen, G. Experimental Investigation of Heat Conduction Mechanisms in Nanofluids. Clue on Clustering. *Nano Lett* **9**, 4128-4132 (2009).
287. Zheng, R. et al. Thermal Percolation in Stable Graphite Suspensions. *Nano Lett* **12**, 188-192 (2011).
288. Zheng, R., Gao, J., Wang, J. & Chen, G. Reversible temperature regulation of electrical and thermal conductivity using liquid–solid phase transitions. *Nat Commun* **2**, 289 (2011).
289. Khalatnikov, I.M. *Sov. Phys. JETP* **22**, 687 (1952).
290. Swartz, E.T. & Pohl, R.O. Thermal-Boundary Resistance. *Reviews of Modern Physics* **61**, 605-668 (1989).
291. Collins, K.C., Chen, S. & Chen, G. Effects of surface chemistry on thermal conductance at aluminum-diamond interfaces. *Applied Physics Letters* **97** (2010).

292. Hopkins, P.E. et al. Manipulating Thermal Conductance at Metal–Graphene Contacts via Chemical Functionalization. *Nano Lett* **12**, 590-595 (2012).
293. Prasher, R. Acoustic mismatch model for thermal contact resistance of van der Waals contacts. *Applied Physics Letters* **94**, 041905 (2009).
294. Hopkins, P.E., Phinney, L.M., Serrano, J.R. & Beechem, T.E. Effects of surface roughness and oxide layer on the thermal boundary conductance at aluminum/silicon interfaces. *Physical Review B* **82**, 085307 (2010).
295. Hsieh, W.-P., Lyons, A.S., Pop, E., Keblinski, P. & Cahill, D.G. Pressure tuning of the thermal conductance of weak interfaces. *Physical Review B* **84**, 184107 (2011).
296. Ge, Z.B., Cahill, D.G. & Braun, P.V. Thermal conductance of hydrophilic and hydrophobic interfaces. *Physical Review Letters* **96** (2006).
297. Shenogina, N., Godawat, R., Keblinski, P. & Garde, S. How Wetting and Adhesion Affect Thermal Conductance of a Range of Hydrophobic to Hydrophilic Aqueous Interfaces. *Physical Review Letters* **102** (2009).
298. Ge, Z.B., Cahill, D.G. & Braun, P.V. AuPd metal nanoparticles as probes of nanoscale thermal transport in aqueous solution. *Journal of Physical Chemistry B* **108**, 18870-18875 (2004).
299. Schmidt, A.J. et al. Probing the gold nanorod-ligand-solvent interface by plasmonic absorption and thermal decay. *Journal of Physical Chemistry C* **112**, 13320-13323 (2008).
300. Losego, M.D., Moh, L., Arpin, K.A., Cahill, D.G. & Braun, P.V. Interfacial thermal conductance in spun-cast polymer films and polymer brushes. *Applied Physics Letters* **97** (2010).
301. Losego, M.D., Grady, M.E., Sottos, N.R., Cahill, D.G. & Braun, P.V. Effects of chemical bonding on heat transport across interfaces. *Nat Mater* **11**, 502-506 (2012).
302. Planck, M. & Masius, M. The Theory of Heat Radiation (Blakiston, 1914).
303. Joulain, K., Mulet, J.-P., Marquier, F., Carminati, R. & Greffet, J.-J. Surface electromagnetic waves thermally excited: Radiative heat transfer, coherence properties and Casimir forces revisited in the near field. *Surface Science Reports* **57**, 59-112 (2005).
304. Narayanaswamy, A. & Chen, G. DIRECT COMPUTATION OF THERMAL EMISSION FROM NANOSTRUCTURES. **14**, 169-195 (2005).
305. Domoto, G.A., Boehm, R.F. & Tien, C.L. Experimental Investigation of Radiative Transfer Between Metallic Surfaces at Cryogenic Temperatures. *Journal of Heat Transfer* **92**, 412-416 (1970).
306. Hargreaves, C.M. Radiative transfer between closely spaced bodies. *Philips Res. Rep. Suppl.* **5**, 1-80 (1973).
307. Kralik, T., Hanzelka, P., Musilova, V., Srnka, A. & Zobac, M. Cryogenic apparatus for study of near-field heat transfer. *Review of Scientific Instruments* **82**, 055106 (2011).
308. Xu, J.-B., Lauger, K., Moller, R., Dransfeld, K. & Wilson, I.H. Heat transfer between two metallic surfaces at small distances. *Journal of Applied Physics* **76**, 7209-7216 (1994).
309. Kittel, A. et al. Near-Field Heat Transfer in a Scanning Thermal Microscope. *Physical Review Letters* **95** (2005).
310. Hu, L., Narayanaswamy, A., Chen, X. & Chen, G. Near-field thermal radiation between two closely spaced glass plates exceeding Planck's blackbody radiation law. *Applied Physics Letters* **92**, 133106 (2008).
311. Ottens, R.S. et al. Near-Field Radiative Heat Transfer between Macroscopic Planar Surfaces. *Physical Review Letters* **107**, 014301 (2011).
312. Narayanaswamy, A., Shen, S. & Chen, G. Near-field radiative heat transfer between a sphere and a substrate. *Physical Review B* **78**, 115303 (2008).
313. Varesi, J., Lai, J., Perazzo, T., Shi, Z. & Majumdar, A. Photothermal measurements at picowatt resolution using uncooled micro-optomechanical sensors. *Applied Physics Letters* **71**, 306-308 (1997).
314. Shen, S., Narayanaswamy, A., Goh, S. & Chen, G. Thermal conductance of bimaterial microcantilevers. *Applied Physics Letters* **92**, 063509 (2008).
315. Gu, N., Sasihithlu, K. & Narayanaswamy, A. JWE13 (Optical Society of America, 2011).
316. Shen, S., Mavrokefalos, A., Sambegoro, P. & Chen, G. Nanoscale thermal radiation between two gold surfaces. *Applied Physics Letters* **100**, 233114 (2012).
317. Rousseau, E. et al. Radiative heat transfer at the nanoscale. *Nat Photon* **3**, 514-517 (2009).
318. Otey, C. & Fan, S. Numerically exact calculation of electromagnetic heat transfer between a dielectric sphere and plate. *Physical Review B* **84**, 245431 (2011).